\documentclass[12pt]{iopart}
\usepackage{graphicx}

\newcommand{\be}{\begin{equation}}
\newcommand{\ee}{\end{equation}}
\newcommand{\bea}{\begin{eqnarray}}
\newcommand{\eea}{\end{eqnarray}}
\newcommand{\beastar}{\begin{eqnarray*}}
\newcommand{\eeastar}{\end{eqnarray*}}

\newcommand{\lav}{\left\langle}
\newcommand{\rav}{\right\rangle}

\newcommand{\half}{\frac{1}{2}}

\newcommand{\eq}[1]{~(\ref{#1})}

\newcommand{\sgn}{\mbox{sgn}}

\newcommand{\order}{{{\mathcal O}}}

\newcommand{\ie}{{\it i.e.}}
\newcommand{\eg}{{\it e.g.}}
\newcommand{\G}{G}
\newcommand{\z}{z}
\newcommand{\Ss}{S}
\newcommand{\n}{\rho}
\newcommand{\N}{N}
\newcommand{\M}{M}
\newcommand{\m}{m}
\renewcommand{\H}{H}
\newcommand{\I}{I}
\newcommand{\Me}{\even{\M}}
\newcommand{\Mo}{\odd{\M}}
\newcommand{\leven}{\even{\lambda}}
\newcommand{\lodd}{\odd{\lambda}}
\newcommand{\sig}{\tau}
\newcommand{\T}{T}
\newcommand{\Ta}{\T_1}
\newcommand{\Tb}{\T_2}
\renewcommand{\ns}{t}
\newcommand{\nsa}{\ns_1}
\newcommand{\nsb}{\ns_2}
\newcommand{\nse}{\even{\ns}}
\newcommand{\nso}{\odd{\ns}}
\renewcommand{\u}{v}
\newcommand{\w}{u}
\newcommand{\entropy}[1]{{\mathcal H}\left(#1\right)}

\renewcommand{\Tr}{{\rm Tr_{\even{\sig},\odd{\sig}}\,}}
\newcommand{\ddx}{\partial_x}
\renewcommand{\c}{_{\rm c}}
\renewcommand{\l}{^{(l)}}
\renewcommand{\lo}{^{(l+1)}}
\newcommand{\even}[1]{#1}
\newcommand{\odd}[1]{{\bar #1}}
\renewcommand{\i}{_-}
\renewcommand{\a}{_+}
\newcommand{\Mmax}{M_{\mathrm{max}}}
\newcommand{\binom}[2]{\left(\!\!\begin{array}{c}#1 \\ #2\end{array}\!\!\right)}

\begin{document}

\title[Phase Transition in a Random Minima Model]{Phase Transition in a Random Minima Model: Mean Field Theory and Exact Solution on 
the Bethe Lattice}

\author{Peter Sollich$^1$, Satya N.\ Majumdar$^2$, Alan J.\ Bray$^3$}

\address{$^1$ Department of Mathematics, King's College London, London WC2R
2LS, UK}
\address{$^2$ Laboratoire de Physique Th\'eorique et Mod\`eles Statistiques,
Universit\'e de Paris-Sud, 91405 Orsay Cedex, France}
\address{$^3$ School of Physics and Astronomy, University of Manchester, 
Manchester M13 9PL, UK}
\ead{peter.sollich@kcl.ac.uk}

\begin{abstract}
  We consider the number and distribution of minima in random
  landscapes defined on non-Euclidean lattices. Using an ensemble
  where random landscapes are reweighted by a fugacity factor $z$ for
  each minimum they contain, we construct first a `two-box' mean field
  theory. This exhibits an ordering phase transition at $z\c=2$ above
  which one box contains an extensive number of minima.  The onset of
  order is governed by an unusual order parameter exponent $\beta=1$,
  motivating us to study the same model on the Bethe lattice.  Here we
  find from an exact solution that for any connectivity $\mu+1>2$
  there is an ordering transition with a conventional mean field order
  parameter exponent $\beta=1/2$, but with the region where this
  behaviour is observable shrinking in size as $1/\mu$ in the mean
  field limit of large $\mu$. We show that the behaviour in the
  transition region can also be understood directly within a mean
  field approach, by making the assignment of minima `soft'. Finally
  we demonstrate, in the simplest mean field case, how the analysis
  can be generalized to include both maxima and minima. In this case
  an additional first order phase transition appears, to a landscape
  in which essentially all sites are either minima or maxima.
\end{abstract}




\section{Introduction}

The statistics of the number of stationary points (maxima, minima and saddles) in a random landscape
plays an important role in understanding both the static and the dynamical properties
of many systems such as structural glasses~\cite{BarratFeigelman02}, spin 
glasses~\cite{MezardParisi87b},
clusters and biomolecules~\cite{Wales}, 
continuum percolation~\cite{WH}, rugged landscapes in evolutionary
biology~\cite{Gavrilets04}, quantum cosmology~\cite{Mersini-Houghton05}, string 
theory~\cite{Susskind03, AazamiEasther06}, particles in a random 
potential~\cite{CavagnaGarrahan99,Fyodorov04}, and also 
in several associated problems in random 
matrices~\cite{AazamiEasther06,CavagnaGarrahan00,DeanMajumdar06,FW}. 
In particular, the statistics of the total number of minima is 
important to understand in the context of glassy materials where the
system typically gets trapped for a
long time in a local 
minimum of the energy (or free energy) landscape~\cite{BarratFeigelman02,MezardParisi87b}. 

In theoretical studies one typically models a random energy landcsape as a smooth
random manifold (typically Gaussian) sitting on an underlying continuous space.
For such smooth Gaussian random surfaces in a continuum, there is a finite density of
local minima (expected number of local minima per unit volume) which can
be computed from the celebrated Kac-Rice formula~\cite{KR} and its multi-dimensional
generalizations~\cite{Belyaev,Cline,Adler}. Recently these formulae have been
extended to compute the expected number of saddle points at a given fixed energy
and also with a fixed index number of saddles~\cite{BD,FSW,VM}.
Similarly the variance and higher moments can also be computed in principle.
However, this machinery is not easily extendable to cases (i) where the 
surface is non-Gaussian and (ii) where the underlying space 
on which the energy landscape resides is discrete, \eg\ a 
regular Euclidean lattice. The latter case is particularly relevant 
in the practical context of numerical simulations of the random energy
landscape where one is obliged to discretize the underlying space.  
Hence it is interesting to study the statistics of the number of
local minima in lattice models of energy landscapes, in particular
where the energy distribution at each site is non-Gaussian in general. 

With these two motivations in mind, a simple lattice model of an energy
landscape has recently been introduced (hereafter referred to as the `random minima' model)~\cite{MM}.
In this random minima model, a random energy 
$E_i$ sits at site $i$ 
of a lattice of $N$ sites with periodic boundary conditions. The energies $E_i$ are drawn,  
independently
from site to site, from a common continuous distribution $p(E)$, not necessarily Gaussian. Any such 
choice of the set $\{E_i\}$ defines a realization of a random energy landscape.
For a given realization, a site $i$ is a local minimum
if $E_i<E_j$ for all sites $j$ which are nearest neighbours of site $i$.
Let $M$ denote the total number of local minima. Evidently $M$
will vary from one realization of the landscape to another and one
is interested in the probability distribution $P(M,N)$ of $M$
for a given size $N$ of the system. 
The energies at different sites are uncorrelated in the random minima model.
Hence it
can be viewed as an effective `coarse grained' lattice model 
of a continuous random energy manifold in the limiting situation 
where the correlation length between the energies at
different points in space is smaller than the lattice spacing.

The advantage of this simple model is that many questions
regarding the statistics of the number of minima are analytically
tractable~\cite{MM}. Besides, the same distribution $P(M,N)$ has appeared recently in seemingly unrelated problems such
as random permutations~\cite{Stembridge97,OshaninVoituriez04},
ballistic deposition models~\cite{HivertNechaev05} and also simple models of 
glasses~\cite{BurdaKrzywicki06}.
The distribution $P(M,N)$ turns out to be strictly universal,
in the sense of being (even for finite $N$) independent of the on-site energy distribution $p(E)$ as long
as the latter is continuous~\cite{MM,HivertNechaev05}. This is not
difficult to see: transform from the $E_i$ to new variables
$x_i=q(E_i)$, with $q(E)=\int_{-\infty}^E dE'p(E')$ the cumulative
distribution function. For continuous $p(E)$, this transformation is
monotonic so that minima in the $E_i$-landscape are identical with
minima in the landscape defined by the $x_i$. But the $x_i$ are
uniformly distributed in the interval $[0,1]$, so it suffices to
consider this particular distribution -- denoted $Q(x)$ below -- to
obtain $P(M,N)$ for {\em any} continuous $p(E)$.

The average number of minima can 
be trivially computed, $\langle M\rangle= N/[(\mu+1)+1]$ where $\mu+1$
is the
co-ordination number of the lattice. For
example, for a $d$-dimensional hypercubic lattice, $\mu=2d-1$ whereas
for a Bethe lattice $\mu$ is just the branching ratio.
Similarly, the variance 
of $M$ can also be computed exactly for various lattices such as
a $1$-d chain~\cite{MM,HivertNechaev05}, the $2$-d square lattice~\cite{HivertNechaev05} and the Bethe 
lattice~\cite{MM}. The distribution
$P(M,N)$ has a Gaussian peak near its mean (of width $\sim \sqrt{N}$),
but a non-Gaussian tail far from the mean. The non-Gaussian tail
is described by a large deviation function that can be computed exactly
in $1$-d~\cite{MM}. Also, on any given lattice, $M$ can at most take a value $\Mmax$.
This follows from the fact that if a site is a local minimum, none of its neighbours
can be a local minimum (nearest neighbour minima exclusion principle). For example,
on a bipartite lattice, consisting of two `boxes' each containing $N$
sites and where each site has nearest neighbour connection to all sites in the other
box, one cannot have minima in both boxes and so $\Mmax=N$. In
Ref.~\cite{MM}, the probability of the maximal packing configuration 
$P(\Mmax,N)$ was studied and was shown
to decay for large $N$ as $P(\Mmax,N)\sim \gamma^{-N}$, where
the constant $\gamma$ was exactly computed for a number of lattices.    

The purpose of this paper is to go beyond the `counting problem' in the
random minima model and study its thermodynamics and the associated phase transition 
by introducing a 
fugacity
$z$ for each local minimum. For this purpose, the relevant object of interest
is the generating function (or grand partition function)
\begin{equation}
G(z,N)= \sum_{M} P(M,N)\, z^M
\label{grandpfminima}
\end{equation}
and the associated equation of state. The latter tells us how the
density of minima $\n=\langle M\rangle/N$ depends on $z$, where the
average is over the original ensemble of random landscapes reweighted
by a factor $z^M$ for each configuration.

At this point it is important to note
that due to the nearest neighbour minima exclusion principle, the thermodynamics
of the random minima model is similar in spirit, though not in its details, to the
well studied `hard sphere lattice gas' model~\cite{Burley,Temperley,GF,RC,Ree-Chesnut,Runnels}. 
In the latter model, 
when a molecule
or a hard sphere occupies a lattice site, a similar exclusion principle holds
in that all neighbouring sites have to be empty. 
If $W(M,N)$ denotes the number of ways of putting $M$ hard particles (with
this constraint of nearest neighbour exclusion) on a lattice of $N$ sites, the
corresponding grand partition function is defined as 
\begin{equation}
Z(z,N)= \sum_{M} W(M,N)\, z^M.
\label{grandpfhs}
\end{equation}
Note the important difference between the two models. In the hard sphere
model with $M$ particles, one attaches a uniform weight $1$ to 
each allowed configuration of the $M$ particles. On the other hand, in
the random minima model, for each configuration of $M$ local minima, the associated
weight comes from an entropic factor obtained by integrating over all possible
$E_i$'s associated with the given configuration of the $M$ local minima.

The hard sphere model is well known to undergo a thermodynamic phase transition
as one increases the fugacity $z$ through a critical value $z_c$ in two or
higher dimensions~\cite{GF,Runnels}. For $z<z_c$,
the system is in a low density `disordered' or `fluid' phase and for $z>z_c$
it is in a high density `ordered' or `crystalline' phase. Based on the
qualitative analogy between the
two models one therefore 
expects
a similar phase transition from a disordered to an ordered state in the random minima 
model, also upon increasing the fugacity $z$. Indeed, recent numerical studies by
Derrida for a $2$-d random minima model indicate the presence of such a phase transition~\cite{Derrida}.
It is important to understand whether this phase transition in the random minima model
is similar/different from that of the hard sphere lattice gas model.

With this in mind, we study the thermodynamics of the random minima model
on the Bethe lattice. The hard sphere model was solved exactly on the
Bethe lattice many years back~\cite{Runnels} and has been revisited recently~\cite{Weigt-Hartmann}. In 
this paper we present an exact solution of the random minima    
model on the Bethe lattice which turns out to be technically somewhat harder
than the hard sphere solution on the same lattice. 
In addition, we study analytically a rather simple mean field theory
of the random minima model which also exhibits a phase transition
at a critical value $z\c=2$. We show that in the low density phase (for $z<2$) the
average number of minima is of order unity in the thermodynamic limit, while
in the high density phase (for $z>2$) the average number of minima is extensive with
a finite density.

We survey the hard sphere lattice gas briefly in
Sec.~\ref{sec:HS}. There is no non-trivial mean field theory for hard
particles, so we start directly with the Bethe lattice case
(Sec.~\ref{sec:HS_Bethe}). A mean field theory {\em can} be constructed if
particles are made soft, i.e.\ if occupation of neighbouring sites is
permitted subject to some penalty. As we show in Sec.~\ref{sec:HS_soft},
when the penalty parameter is made large this approach nicely retrieves the
results for the Bethe lattice in the limit of large connectivity.

We turn to our main subject, the random minima problem, in
Sec.~\ref{sec:minima}. Here there is a non-trivial two-box mean field theory
and we discuss this first, in Sec.~\ref{hard minima}, and also extend
it to study the joint statistics of the number of minima and maxima
(Sec.~\ref{sec:minima_maxima}). Next we analyse the random minima
problem on the Bethe lattice (Sec.~\ref{sec:Bethe_min}) and finally we
consider a mean field theory with soft assignments of minima in
Sec.~\ref{sec:Bethe_soft}. Again we will see that these two approaches
give the same results in the respective limits of large connectivity and
almost-hard assignments. Considering these limit cases also helps to
clarify why the direct mean field approach of Sec.~\ref{hard minima}
gives an unusual apparent order parameter exponent near the phase
transition. We summarize and list some open questions in
Sec.~\ref{sec:summary}.

\section{Hard sphere lattice gas}
\label{sec:HS}

In this section we revisit briefly the hard sphere lattice gas model,
on a Bethe lattice~\cite{Runnels,Weigt-Hartmann} and in a two-box
mean field theory with soft particles. This will serve to introduce the techniques we will
deploy later for the problem of
minima in random landscapes. Note that the simplest mean field theory,
a fully connected lattice, makes no sense as the presence of a
single particle would exclude particles from all other sites. Also the
simplest improvement over this, a fully connected bipartite lattice,
is trivial: one of the two boxes, i.e.\ partitions of the graphs, is
always empty, and in the other the particles are then non-interacting. In the
random minima problem, on the other hand, already this approach
produces a phase transition as discussed in Sec.~\ref{hard minima} below.

\subsection{Bethe lattice}
\label{sec:HS_Bethe}

Consider first a Cayley tree\footnote{%
We note that in some of the literature what we call a Cayley tree is
termed ``rooted Cayley tree'', while the term ``Cayley tree'' is then
used for what we call a Bethe lattice, i.e.\ a tree where every node
except those on the boundary has $\mu+1$ neighbours.
}
with branching ratio $\mu$, of depth $l$,
\ie\ with $l$ layers below the single root node. Call $Z\l_{0,1}$ the
grand partition function constrained to run over all configurations
that do not (or do, respectively) have a particle at the root. The
full partition function is them $Z\l=Z\l_0+Z\l_1$. We do not write
explicitly the dependence on $z$, while the superscript $(l)$
indicates indirectly the number of sites $N=(\mu^{l+1}-1)/(\mu-1)$ in
the tree.

The quantities $Z\l_{0,1}$
obey the following recursions over the tree depth:
\bea
Z\lo_0 &=& (Z\l_0+Z\l_1)^\mu
\\
Z\lo_1 &=& \z(Z\l_0)^\mu
\eea
with $Z^{(0)}_0=1$, $Z^{(0)}_1=\z$. For example, if no particle is
present at the root node of a tree of depth $l+1$, then the $\mu$
sites in the next level of the tree are each allowed to be either
occupied or not; the partition sum is then the product of $\mu$
unconstrained partition functions $Z\l_0+Z\l_1$ for each of the
subtrees of depth $l$. This gives the first equation above. For the
second equation, one notes that if a particle is present at the root
then each of the $\mu$ sites below must be empty. The partition sum is
then a product of the appropriate constrained partition sums $Z\l_0$ for the
subtrees, with an extra factor $\z$ to account for the particle at the
root.

\begin{figure}
\begin{center}
\includegraphics[width=6cm]{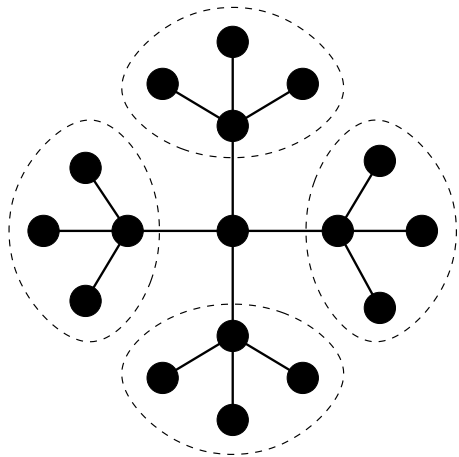}
\end{center}
\caption{Sketch of a Bethe lattice with $\mu=3$: every interior node
  has $\mu+1=4$ neighbours. This lattice can be obtained by connecting
  $\mu+1=4$ Cayley trees -- indicated by the dashed lines -- of
  branching ratio $\mu=3$ (and in this case depth $l=1$) to the
  central node.
\label{fig:bethe}
}
\end{figure}
The two recursions can be combined into one for the ratio
$S\l=Z\l_1/Z\l_0$, giving
\be
\Ss\lo = \frac{\z}{(1+\Ss\l)^\mu}
\label{S_recursion}
\ee
with $\Ss^{(0)}=\z$. From $\Ss\l$ one can determine the density at the
centre of a Bethe lattice of depth $l+1$, obtained by connecting
$\mu+1$ Cayley trees of depth $l$ to a central node (see
Fig.~\ref{fig:bethe}). Taking the
appropriate ratio of the partition sum with the central site occupied
to the total partition sum yields
\bea
\n\lo &=&
\frac{\z(Z\l_0)^{\mu+1}}{(Z\l_0+Z\l_1)^{\mu+1}+\z(Z\l_0)^{\mu+1}}
\\
&=& \frac{\z}{(1+\Ss\l)^{\mu+1}+\z}\ .
\label{hard_density}
\eea
For low $\z$ the recursion for $\Ss\l$ has a single fixed point. The
bifurcation to an ordered state occurs when
$\partial\Ss\lo/\partial\Ss\l=-1$ at the fixed point\footnote{%
If we write the recursion as $\Ss\lo=f(\Ss\l)$, then the bifurcation is
to a cycle of two solutions, $\even{\Ss}=f(\odd{\Ss})$ and
$\odd{S}=f(\even{\Ss})$. Near the bifurcation $\even{\Ss}$ and $\odd{S}$
are close, so one can expand
$\even{\Ss}=f(\even{\Ss})+(\odd{\Ss}-\even{\Ss})f'(\even{\Ss})
+O((\odd{\Ss}-\even{\Ss})^2)$ which gives
$(\even{\Ss}-\odd{\Ss})[1+f'(\even{\Ss})]=O((\odd{\Ss}-\even{\Ss})^2)$ and
hence $f'(\even{\Ss})=-1$ at the bifurcation itself.%
}. This
requirement together with the fixed point condition itself gives
\be
\Ss\c=\frac{1}{\mu-1}, \qquad
\z\c=\mu^\mu/(\mu-1)^{\mu+1}, \qquad
\n\c=\frac{1}{\mu+1}\ .
\label{hard_critical_point}
\ee
The divergence of
$\z\c$ at $\mu=1$ makes sense: for $\mu=1$ we have a chain, which as a
one-dimensional system with only short-range interactions cannot
exhibit a phase transition.

For $\z>\z\c$ the recursion for $\Ss\l$ converges to a period-two
sequence, $\Ss^{(2k)}\to\even{\Ss}$, $\Ss^{(2k+1)}\to\odd{\Ss}$, where
\be
\even{\Ss} = \frac{\z}{(1+\odd{\Ss})^\mu}, \qquad
\odd{\Ss} = \frac{\z}{(1+\even{\Ss})^\mu}\ .
\label{hard_ordered_fixed_pt}
\ee
As anticipated, this means the system is ordered, with alternating
layers of the lattice preferentially occupied/empty; the densities
are:
\bea
\even{\n} &=& \frac{\z}{(1+\odd{\Ss})^{\mu+1}+\z} = 
\frac{1}{\z^{1/\mu}\even{\Ss}^{-(\mu+1)/\mu}+1}
\\
\odd{\n} &=& \frac{\z}{(1+\even{\Ss})^{\mu+1}+\z} = 
\frac{1}{\z^{1/\mu}\odd{\Ss}^{-(\mu+1)/\mu}+1}\ .
\eea
(Given the initial condition $\Ss^{(0)}=\z>\Ss\c$, the even layers
should be the occupied ones, \ie\ $\even{\Ss}>\odd{\Ss}$.)
Mathematically, further bifurcations could occur for larger $\z$, but
physically this is implausible. In general, the equations for
$\even{\Ss}$ and $\odd{\Ss}$ need to be solved numerically. For large
$\mu$ simplifications occur, however. It is to this mean field limit
that we now turn.

\subsubsection{Large $\mu$, above the transition}

Here we take fixed $\z>\z\c$. It is then not hard to see that for
$\mu\to\infty$ one gets $\even{\Ss}=\z$,
$\odd{\Ss}=\z(1+\z)^{-\mu}$. (This is self-consistent, 
since $\odd{\Ss}$ is exponentially small and so $(1+\odd{\Ss})^\mu\to 1$.)
The resulting densities are, to leading order,
\be
\even{\n} = 1/(\z^{-1}+1)=\z/(\z+1), \qquad
\odd{\n}=\z(1+\z)^{-(\mu+1)}\ .
\label{n_hard_large_mu}
\ee
This is plausible: $\odd{\n}$ is very small so that the odd layers are
basically empty. Then $\even{\n}$ is just determined by the activity
$\z$, which attributes weights $\z$ and $1$, respectively, to the
configuration with or without a particle on a site of the even sublattice.

\subsubsection{Large $\mu$, around the transition}

For $\mu\to\infty$, the critical activity
from\eq{hard_critical_point} becomes $\z\c=e/\mu$. We therefore set
$\z=\tilde\z/\mu$ to explore the region around the ordering
transition. Since the critical 
density $\n\c=1/(\mu+1)$ and partition sum ratio $\Ss\c=1/(\mu-1)$ are also
$\order(1/\mu)$, we put likewise $\n=\tilde\n/\mu$ and
$\Ss=\tilde\Ss/\mu$. The fixed point in the disordered phase then obeys,
from the large-$\mu$ limit of the fixed point of\eq{S_recursion},
\be
\tilde\Ss=\tilde\z e^{-\tilde\Ss}
\ee
From\eq{hard_density} the density becomes $\tilde\n=\tilde\z
e^{-\tilde\Ss}=\tilde\Ss$, so the 
`equation of state' is simply
\be
\tilde\z=\tilde\n e^{\tilde\n}\ .
\ee
In the ordered phase, on the other hand, one has from\eq{hard_ordered_fixed_pt}
\be
\tilde\even{\Ss} = \tilde\z e^{-\tilde\odd{\Ss}}, \qquad
\tilde\odd{\Ss} = \tilde\z e^{-\tilde\even{\Ss}}
\ee
with again $\tilde\even{\n} = \tilde\z e^{-\tilde\odd{\Ss}}=\tilde\even{\Ss}$
and $\tilde\odd{\n} = \tilde\odd{\Ss}$. So the activity and the densities
in the even/odd layers are related by
\be
\tilde\z = \tilde\even{\n} e^{\tilde\odd{\n}} = \tilde\odd{\n}
e^{\tilde\even{\n}} \ .
\ee
The two densities obey
\be
\tilde\even{\n} e^{-\tilde\even{\n}}=\tilde\odd{\n} e^{-\tilde\odd{\n}}
\label{n_hard_two_densities}
\ee
and hence the critical point is at
$\tilde\even{\n}=\tilde\odd{\n}=\tilde\n\c=1$, $\tilde\z\c=e$ as
expected. For higher $\tilde\z$, the densities deviate from each other
with a standard square root singularity, to leading order,
$\tilde\even{\n}-1=1-\tilde\odd{\n}\sim (\tilde\z-\tilde\z\c)^{1/2}$.
This corresponds to an order parameter critical exponent $\beta=1/2$
as expected for a mean field model.
For general $\tilde\z>\tilde\z\c$, the last two equations -- which
together determine the equation of state of the ordered phase -- need
to be solved numerically. (One could choose, say, $\tilde\even{\n}>1$,
find the corresponding $\tilde\odd{\n}$ from\eq{n_hard_two_densities},
then determine $\tilde\z$.) The asymptotic behaviour for $\tilde\z \gg
1$ is $\tilde\even{\n}=\tilde\z$, $\tilde\odd{\n}=\tilde\z
e^{-\tilde\z}$ which matches with the $\z\ll 1$ limit
of\eq{n_hard_large_mu} as it should.

\subsection{Mean field theory with soft particles}
\label{sec:HS_soft}

The large connecitivity limit $\mu\to\infty$ discussed above must
correspond to a mean field theory that one ought to be able to construct directly,
without having to first solve for lattices of finite connectivity.
As explained above, a fully connected lattice makes no sense as the presence of a
single particle would exclude particles from all other sites.  One is
therefore led to considering a fully connected bipartite lattice. This
can be thought of as two boxes (`left' and `right') with $\N$
sites each; every site is connected to all others in the {\em other}
box. If we now directly enforce the hard repulsion of particles on
neighbouring (connected) sites, the model is trivial: as soon as one
box contains any particles, the other one must be completely
empty. The density in the non-empty box is then just
$\even{\n}=\z/(\z+1)$ as determined by the fugacity, and the system is
always ordered.

To retrieve the ordering phase transition, one needs to
introduce a soft repulsion. Here we give a configuration with
$\even{\M}$ and $\odd{\M}$ particles in the two boxes weight
$\z^{\even{\M}+\odd{\M}}\exp(-\alpha \even{\M} \odd{\M}/\N)$. Sending
$\alpha\to\infty$ then recovers the hard repulsion.

The partition function for this soft repulsion model is
\be
Z(z,N)=\sum_{\even{\M},\odd{\M}}
\binom{\N}{\even{\M}}\binom{\N}{\odd{\M}}
\z^{\even{\M}+\odd{\M}}e^{-\alpha \even{\M} \odd{\M} / \N}
\ee
and can be evaluated by introducing the densities
$\even{\n}=\even{\M}/\N$, $\odd{\n}=\odd{\M}/\N$ and evaluating using
steepest descents for $\N\to\infty$:
\be
\N^{-1}\ln Z = \max_{\even{\n},\odd{\n}} \left\{
\entropy{\even{\n}}+\entropy{\odd{\n}}+
(\even{\n}+\odd{\n})\ln\z-\alpha \even{\n} \odd{\n}\right\}
\ee
with the entropy
\be
\entropy{\even{\n}}=-\even{\n}\ln\even{\n} - (1-\even{\n})\ln(1-\even{\n})\ .
\ee
The resulting saddle point conditions are
\bea
\ln[(1-\even{\n})/\even{\n}] + \ln\z - \alpha \odd{\n} &=& 0
\\
\ln[(1-\odd{\n})/\odd{\n}] + \ln\z - \alpha \even{\n} &=& 0\ .
\eea
We will be interested in the large $\alpha$ limit where the repulsion
is `nearly hard'; when the symmetry between boxes is broken, we
assume without loss of generality that it is the left box that has the
higher density, \ie\ $\even{\n}>\odd{\n}$.

\subsubsection{Large $\alpha$, above the transition}

Taking $\alpha$ large at fixed $\z$, we see that to satisfy the second
saddle point equation to $\order(\alpha)$ one needs
$\odd{\n}=\z\exp(-\alpha \even{\n})$ to leading order. The
$\alpha$-dependent term in the first saddle point equation then
becomes negligible, so that $(1-\even{\n})/\even{\n}=1/\z$ or
$\even{\n}=\z/(\z+1)$. This is just the result\eq{n_hard_large_mu} on
the Bethe lattice for large $\mu$, as expected, and is consistent with
the simple expression obtained from the balance of
the weights of unoccupied and occupied configurations (see above). Note that the
density of the almost empty box is $\odd{\n}=\z\exp[-\alpha\z/(\z+1)]$
to leading order; this {\em does not} match with the Bethe lattice
result if one naively identifies $\alpha$ with $\mu$. So only the
leading order densities ($\z/(\z+1)$ and $0$) match while the
subleading (exponentially small, in the nearly empty box) corrections
are not related.

\subsubsection{Large $\alpha$, around the transition}

Here we set $\even{\n}=\tilde\even{\n}/\alpha$ and
$\odd{\n}=\tilde\odd{\n}/\alpha$, by analogy with the large $\mu$
treatment on the Bethe lattice. This gives for $\alpha\to\infty$ the
saddle point equations
\be
\ln(\tilde\z/\tilde\even{\n}) - \tilde\odd{\n}  = 0, \qquad
\ln(\tilde\z/\tilde\odd{\n})  - \tilde\even{\n} = 0\ .
\ee
These can be rewritten as
\be
\tilde\z=\tilde\even{\n} e^{\tilde\odd{\n}} = \tilde\odd{\n} e^{\tilde\even{\n}}
\ee
which is {\em exactly} the same as the large-$\mu$ result on the Bethe
lattice. So near their respective transitions the two models behave
identically, demonstrating that the fully connected two-box
model with soft repulsion captures the same physics as the Bethe
lattice for high connectivity.

\section{Random minima}
\label{sec:minima}

In this section we turn to our main subject, the arrangement of the
local minima of a random function on a Bethe lattice. As explained in the
introduction, we can without loss of generality take the function
value at each site $i$ to be a random variable $x_i$ sampled from a
uniform distribution $Q(x)$ over $[0,1]$. We will define binary indicator
variables $m_i$, setting $\m_i=1$ if site $i$ is a minimum, \ie\ if none of its neighbours has
a larger $x$; otherwise we set $\m_i=0$. (With this convention, leaves of
a tree are counted as minima if their $x$ is smaller than that of the
parent node directly above.) The $\m_i$ are analogous to hard particle
occupation numbers since no two neighbouring nodes can have $\m=1$;
but the random values $x_i$ introduce other, non-trivial
correlations. As in the hard sphere model we will multiply the
weight of any
configuration of the $\m_i$, produced by a random draw of the $x_i$,
by a fugacity factor $z^\M$, where now $\M=\sum_i \m_i$ is the total number of
minima. We wish to calculate the density of minima as a function of
$\z$, and understand whether an ordering transition does again take
place for sufficiently large $\z$.

\subsection{Mean field theory}
\label{hard minima}

We begin with the simplest calculation, which is the two-box mean
field theory. Each box contains $N$ sites as before,
and the particles in each  box are regarded as
neighbors of all the particles in  the other box. Clearly, all sites
that are minima must belong to the same box. Call the random variables
in the left box $\even{x}_i$ and those in the right box $\odd{x}_i$,
with $i=1,\ldots,N$.

To work out the generating function\eq{grandpfminima} we need to find
$P(M,N)$, the probability of having $M$ minima in our system. Assume
first that the minima are in the left box. There are $M$ minima if
precisely $M$ among the $\even{x}_i$ are 
smaller than all of the $\odd{x}_j$, \ie\ smaller than 
$\odd{x}\i$, the smallest of the $\odd{x}_j$.
Given that the $\odd{x}_j$ are uniformly distributed over $[0,1]$, the
probability distribution of $\odd{x}\i$ is
\begin{equation}
P(\odd{x}\i) = N(1-\odd{x}\i)^{N-1}\ ,
\end{equation}
so the probability that $M$ of the $\even{x}_i$'s are smaller than $\odd{x}\i$ is
\begin{equation}
\binom{N}{M}
\left\langle \odd{x}\i^M (1-\odd{x}\i)^{N-M}\right\rangle
\end{equation}
where the average is taken using the distribution
$P(\odd{x}\i)$. Accounting for the configurations where the roles of
the two boxes are swapped, the probability of having $M$ minima is
twice as large: 
\begin{eqnarray}
P(M,N) & = & 2N \binom{N}{M}
\int_0^1 d\odd{x}\i\,\odd{x}\i^M (1-\odd{x}\i)^{2N-M-1}
\label{inter}
\\
& = & \frac{N!(2N-M-1)!}{(N-M)!(2N-1)!}\ .
\label{pMN}
\end{eqnarray}
Note that the value $M=0$ is impossible as there is always at least
one minimum present, and accordingly one has the normalization
$\sum_{M=1}^N P(M,N)=1$ as is easily checked.

There is in fact a simple counting argument that leads directly to the
result\eq{pMN}. To generate the $N$ random numbers in each box, we can
first sample $2N$ random numbers $y_1,\ldots,y_{2N}$, again from the
uniform distribution over $[0,1]$. We then take a bag containing $N$
labels `left' and $N$ labels `right' and, for each of the $y_i$, pull
out one label from the bag and put $y_i$ in the relevant box. Because
the order in which we consider the different $y_i$ for labelling is
irrelevant, we can in particular take them to be ordered,
$y_1<\ldots<y_{2N}$. Then a configuration with $M$ minima in the left
box is one where $y_1$ to $y_M$ have got labels `left' and $y_{M+1}$
the label `right'. Keeping track how many `left' and `right' labels
remain in the bag at each step of the labelling, and including the
overall factor of 2 for the reverse situation where the $M$ minima are
in the right box gives
\begin{eqnarray}
P(M,N) &=& 2 \,\frac{N}{2N}\cdot\frac{N-1}{2N-1}\cdots\,\frac{N-(M-1)}{2N-(M-1)}
\ \cdot\ \frac{N}{2N-M}
\\
&=& 2\, \frac{N!}{(N-M)!}\,\frac{(2N-M)!}{(2N)!}\,\frac{N}{2N-M}
\\
& = &
2 
\binom{2N-M-1}{N-1} \binom{2N}{N}^{-1} \ .
\end{eqnarray}
The second expression is the one most easily seen to agree
with\eq{pMN}. The third one gives another way of thinking about the
result: having $M$ minima (in the left box) fixes the first $M+1$
labels, and the probability is then the number of arrangements of the
remaining labels divided by the number of arrangements of {\em all}
labels.

Returning now to our original aim of computing the generating function
$G(z,N)$, it is in fact
most convenient to employ the integral form (\ref{inter}) of
$P(M,N)$. Then the sum over $M$ in Eq.\ (\ref{grandpfminima}) can be
evaluated to give 
\begin{equation}
\fl G(z,N) = 2N\int_0^1
d\odd{x}\i\,(1-\odd{x}\i)^{N-1}
\left\{[1+(z-1)\odd{x}\i]^N-(1-\odd{x}\i)^N\right\}\ . 
\label{genresult}
\end{equation}
The following analysis shows that there is a phase transition, at $z\c=2$, 
between a phase where the number of minima remains finite as $N \to \infty$ 
and a phase where the minima are extensive in number.  
We begin by setting $\odd{x}\i=v/N$ in Eq.\ (\ref{genresult}), and taking the limit 
$N \to \infty$ at fixed $v$. In this limit the integrand becomes 
$\exp[-v(2-z)]-\exp(-2v)$ and the upper limit on $v$ tends to infinity, to give
\begin{equation}
G(z,N\to\infty) = \frac{1}{2-z} - \frac{1}{2} = \frac{z}{2-z},\  \ \ z
< 2 .
\label{GhardMin}
\end{equation}
The same expression can be obtained by noting that, for finite $M$ and large $N$,
the number of minima has the geometric distribution $P(M,N\to\infty) = 2^{-M}$.
The result\eq{GhardMin} diverges at $z =2$. This indicates that the
underlying assumption -- 
that values of $\odd{x}\i$ of order $1/N$ (corresponding to $M$ of order
unity) dominate the integral in (\ref{genresult}) --  
no longer holds, and suggests a phase transition at $z\c=2$. 

For $z>2$, the integral can be evaluated using the method of steepest 
descents. For this purpose we write the integral (\ref{genresult}) in the form
\begin{equation}
G(z,N) = N \int_0^1
\frac{d\odd{x}\i}{1-\odd{x}\i}\{(1-\odd{x}\i)[1+(z-1)\odd{x}\i]\}^N\ .
\end{equation}
We have discarded the second term in the integrand, which is
exponentially subdominant except at $\odd{x}\i=0$. 
For $z>2$, the remaining integral is now dominated by values of $\odd{x}\i$ near the one that 
maximises the function $g(\odd{x}\i) = (1-\odd{x}\i)[1+\odd{x}\i(z-1)]$. This value is 
$x^* = (z-2)/[2(z-1)]$; the fact that $x^*>0$ for $z>2$ justifies a
posteriori why we were able to discard the second term from (\ref{genresult}). Inserting $\odd{x}\i=x^*$
into the integrand now gives 
\begin{equation}
\ln G(z,N\to\infty) = N \ln\left(\frac{z^2}{4(z-1)}\right),\ \ z>2 ,
\end{equation}
up to subextensive contributions. The value $x^*$ has a natural
interpretation: in the $z$-weighted ensemble it is the minimal value
of the random numbers $\odd{x}_i$ in the box without the minima. The
other numbers in this box are then distributed uniformly over
$[x^*,1]$; setting $z>2$ is (for $N\to\infty$) sufficient to exclude
any smaller values.  In the box with the minima, on the other hand,
random numbers from the whole interval $[0,1]$ occur, but the
probability density is higher by a factor $z$ on $[0,x^*]$ than on
$[x^*,1]$ (and uniform within these two intervals). 

We can use the expressions for $G$ in the two regimes to compute the 
expectation value, $\langle M \rangle$, of the number of minima using 
$\langle M \rangle = \sum_M M P(M,N) z^M/\sum_M P(M,N) z^M = 
d\ln G/d\ln z$ to obtain, in the limit $N \to \infty$, 
\begin{equation}
\langle M \rangle = \cases{\frac{2}{2-z},\hspace{1.3cm} z<2, \\
                   N\left(\frac{z-2}{z-1}\right),\ \ z>2.}
\end{equation}

\subsection{Mean field theory: Minima and Maxima}
\label{sec:minima_maxima}

Within the same mean field model, we can also compute the probability 
weights for configurations  which contain maxima as well as minima.
Call the number of minima $M_1$ and the number of maxima $M_2$, and
the probability of such a configuration $P(M_1,M_2,N)$. The most
direct way of obtaining this probability is from the counting argument
outlined above. The result is
\bea
\fl P(M_1,M_2,N) &=& 2 \ \frac{
\binom{2N\!-\!M_1\!-\!M_2\!-\!2}{N-2}
+
\binom{2N\!-\!M_1\!-\!M_2\!-\!2}{N-M_1-1}
+ \delta_{M_1,N}\delta_{M_2,N}}{\binom{2N}{N}}
\label{PM1M2}
\eea
and can be explained as follows. Take the case where the $M_1$
minima are in the left box; the prefactor 2 then accounts for the
opposite case
where they are in the right box. Now the $M_2$ maxima can either be in
the left or the right box. Suppose they are in the left box. Then in
our construction of first drawing $y_1,\ldots,y_{2N}$ and then
labelling them, $y_1,\ldots,y_{M_1}$ and $y_{2N-M_2+1},\ldots,y_{2N}$
need to have label `left' while $y_{M_1+1}$ and $y_{2N-M_2}$ have
label `right'. The number of arranging the $2N-M_1-M_2-2$
labels that remain in the bag, of which $N-2$ are `right', is given by
the first binomial coefficient in the square brackets
in\eq{PM1M2}. The second term is constructed in the same way but with
the labels of $y_{2N-M_2},\ldots,y_{2N}$ reversed: now
$N-M_1-1$ `left' and $N-M_2-1$ `right' labels remain in the bag. This
counting argument works while $M_1\leq N-1$ and $M_2\leq N-1$ (since
we fix $M_1+1$ labels at the bottom and $M_2+1$ labels at the
top). There is only one configuration that is not captured, namely,
$M_1=M_2=N$, where all labels are fixed: the third term 
of\eq{PM1M2} accounts for this.

For finite $M_1$ and $M_2$, where the third term of\eq{PM1M2} is
irrelevant, one easily sees that $P(M_1,M_2,N\to\infty)=2^{-M_1-M_2}$:
the populations of minima and maxima are uncorrelated. (Configurations
with the minima and maxima in the same and in different boxes also have the
same weight, each contributing half the result.) Defining a generating
function
\be
G(z_1,z_2,N)=\sum_{M_1,M_2}P(M_1,M_2,N) z_1^{M_1} z_2^{M_2}\ ,
\label{Gz1z2}
\ee
this implies $G(z_1,z_2,N\to\infty)=[z_1/(2-z_1)][z_2/(2-z_2)]$ for
$z_1<2$ and $z_2<2$. In the ensemble weighted by $z_1$ and $z_2$ the
average numbers of minima and maxima are then $\langle M_1\rangle =
2/(2-z_1)$ and $\langle M_2\rangle = 2/(2-z_2)$, respectively.

\begin{figure}
\begin{center}
\includegraphics[width=6cm]{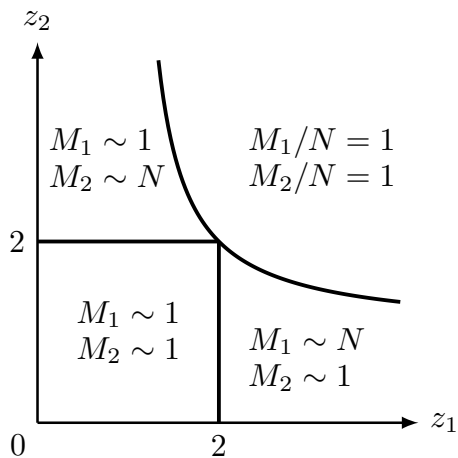}
\end{center}
\caption{Sketch of the mean field phase diagram for the number of
  minima, $M_1$, and maxima, $M_2$, in the random energy landscape
  model.
\label{fig:phasediag}
}
\end{figure}
For larger $z_1$ or $z_2$ one can proceed using steepest descents as
explained in \ref{app:minmax}. We find that there is a first-order transition line
at $1/z_1+1/z_2=1$. Beyond this (\ie\ for $1/z_1+1/z_2<1$), the
numbers in the two boxes separate essentially completely, with one
containing only minima and the other maxima: $\langle
M_1\rangle/N=\langle M_2\rangle/N=1$. Between this transition and the
other boundaries at $z_1=2$ and $z_2=2$ lie two regions where the
number of minima is extensive but the number of maxima is not, and vice
versa. E.g.\ when $z_1>2$ and $z_2<z_1/(z_1-1)$ one finds
\begin{eqnarray}
\langle M_1 \rangle & = & N \left(\frac{z_1-2}{z_1-1}\right)
\ , \\
\langle M_2 \rangle & = & \frac{2(z_1^2+z_2^2)-2z_1z_2(z_1+z_2)+z_1^2z_2^2}
{(2-z_2)(z_1-z_2)(z_1+z_2-z_1z_2)}\ ,
\label{minmax_mixed}
\end{eqnarray}
with an analogous result when the roles of $z_1$ and $z_2$ are
swapped. The last factor in the denominator for $\langle M_2\rangle$
diverges at $1/z_1+1/z_2=1$, signalling the transition to the regime
where both $M_1$ and $M_2$ are extensive. Figure~\ref{fig:phasediag}
shows a sketch of the overall phase diagram. The first order
transition at $1/z_1+1/z_2=1$ has unusual features (see
\ref{app:minmax}): at the transition, an entire one-parameter family
of phases becomes degenerate to leading order, i.e.\ has the same
value of $N^{-1}\ln G$. This should produce unusual finite-size
scaling effects which we have not yet explored.

\subsection{Bethe lattice}
\label{sec:Bethe_min}

Returning to the random minima problem, let us summarize the results so
far. Within a two-box mean field theory, we found that the typical
densities of minima in both boxes, $\even{\n}=\langle \even{M}\rangle/N$ and
$\odd{\n}=\langle \odd{M}\rangle/N$, vanish for fugacities $z<z\c=2$. For
higher fugacities, the system orders, with a nonzero
density of minima $\even{\n}=(z-2)/(z-1)$ in one box but a vanishing
one in the other, $\odd{\n}=0$. A peculiar aspect of this behaviour is
that the nonzero minima density increases {\em linearly} with
$z-z\c=z-2$ around the transition, suggesting an order parameter
exponent $\beta=1$. For a mean field system this would be very unusual
indeed as one would naively expect $\beta=1/2$ in mean field theory. We therefore next consider the minima problem on a Bethe
lattice of finite connectivity. While the large connectivity limit should
then retrieve the mean field results, at finite connectivity we would
hope that a standard mean field phase transition with $\beta=1/2$ will
reappear. This is indeed what we find.

We begin as in the hard particle scenario by considering a Cayley
tree. The basic quantity of 
interest is $P\l(\M\l,\m\l,x\l)$, the probability -- under random
sampling of the $x_i$ -- that the root node of a Cayley tree of depth
$l$ has function value $x\l$, that it is (or is not) a minimum as
indicated by $\m\l=1$ ($\m\l=0$), and that there are a total number $\M\l$ of
minima in the tree. The basic recursion for this is
\bea
\fl 
P\lo(\M\lo,\m\lo,x\lo)&=&Q(x\lo)\prod_{i=1}^\mu\left(\sum_{\M\l_i,\m\l_i}
\int\!dx\l_i P\l(\M\l_i,\m\l_i,x\l_i)\right)\times
\nonumber\\
& &\times \delta_{\M\lo,\ldots} \delta_{\m\lo,\ldots}
\eea
which expresses the fact that $x\lo$ at the new root node is chosen
independently of what happens in the $\mu$ different branches
$i=1,\ldots,\mu$ attached to it. Once $x\lo$ and the properties of
these branches are 
known, the values $\M\lo$ and $\m\lo$ for the new ($l+1$)-level tree
are fully determined as indicated schematically by the
delta-functions. Explicitly,
\bea
\m\lo &=& \left\{\begin{array}{cl}
1 & \mbox{if}\ x\lo < x\l_i\ \forall i=1,\ldots,\mu\\
0 & \mbox{otherwise}\end{array}\right.
\\
\M\lo &=& \M\l + \m\lo - \sum_{i=1}^\mu m\l_i \Theta(x\l_i-x\lo)\ .
\label{M_recursion}
\eea
The last sum runs over the $\mu$ nodes below the new root node as
before. It expresses the fact that even if these nodes were minima
within their own subtrees, once they are connected to the new root
node they cease to be minima if they have function values
$x\l_i>x\lo$.

Introducing the generating functions for $P\l$,
\be
\G\l_{\m}(x) = \sum_{\M=0}^\infty \z^{\M} P\l(\M,\m,x)
\ee
where the fugacity $\z$ again acts on the number of minima, the
recursion becomes (the restriction $0\leq x\leq 1$
is understood for all $x$-variables and so in particular $Q(x\lo)=1$):
\bea
\fl \G\lo_0(x)&=&\left(
\int_0^x\!\!dy\left[\G\l_0(y)+\G\l_1(y)\right]+
\int_x^1\!\!dy\left[\G\l_0(y)+\z^{-1}\G\l_1(y)\right]
\right)^\mu\nonumber\\
& &{}-{}
\left(
\int_x^1\!\!dy\left[\G\l_0(y)+\z^{-1}\G\l_1(y)\right]
\right)^\mu
\label{Gl0_recursion}
\\
\fl \G\lo_1(x)&=&\z\left(
\int_x^1\!\!dy\left[\G\l_0(y)+\z^{-1}\G\l_1(y)\right]
\right)^\mu\ .
\eea
The second of these is easiest to explain: if the root node is a
minimum with function value $x$, all $\mu$ nodes in the level below
must have function values $x\l_i\equiv y>x$. The factor $z^{-1}$ in
front of $G\l_1(y)$ corresponds to the negative term
in\eq{M_recursion}, \ie\ the fact that none of these nodes can then
be minima. The prefactor $\z$ accounts for the new minimum at the root. The
recursion\eq{Gl0_recursion} works similarly: multiplying out the
$\mu$-th power in the first line and subtracting the term in the
second line gives all the possible configurations where at least one
of the nodes below the new root has a lower function value than the
latter. The factor of $\z^{-1}$ is again for nodes which have higher
function values than the new root and so cease to be minima if that is
what they previously were.

As in the hard particle case, a simpler recursion is obtained by
taking ratios of appropriate generating functions. Here, it turns out
to be convenient to consider the ratio
$\Ss\l_\m(x)=\G\l_\m(x)/\G\l_0(1)$. We also abbreviate
\bea
\H\l(x)&=&\int_x^1\!\!dy\left[\Ss\l_0(y)+\z^{-1}\Ss\l_1(y)\right] \\
\I\l(x)&=&\int_0^x\!\!dy\left[\Ss\l_0(y)+\Ss\l_1(y)\right]+
\int_x^1\!\!dy\left[\Ss\l_0(y)+\z^{-1}\Ss\l_1(y)\right]\ .
\eea
Then our recursions read simply:
\bea
(\lambda\l)^\mu\Ss\lo_0(x) &=& \I\l(x)^\mu - \H\l(x)^\mu
\\
(\lambda\l)^\mu\Ss\lo_1(x) &=& \z\H\l(x)^\mu
\eea
where $(\lambda\l)^\mu=\I\l(1)^\mu$ is the normalizing coefficient that
enforces $\Ss\lo_0(1)=1$ for all $l$ as it must be.
The corresponding differential versions will be more useful for
later: expressing the $\Ss\lo_m(x)$ as derivatives of $\H\lo(x)$ and
$\I\lo(x)$ gives
\bea
(\lambda\l)^\mu\ddx\H\lo(x) &=& -\I\l(x)^\mu
\\
(\lambda\l)^\mu\ddx\I\lo(x) &=& (\z-1)\H\l(x)^\mu
\eea
with boundary conditions
\be
\H\l(1)=0, \qquad
\H\l(0)=\I\l(0)
\label{boundary_H_I}\ .
\ee
Once we have the functions $\H\l(x)$ and $\I\l(x)$, the density (\ie\
the probability of having a minimum) at the central node of a Bethe
lattice follows directly as
\be
\n\lo = \frac{\z\int dx\,\H\l(x)^{\mu+1}}
{\int dx\,[\I\l(x)^{\mu+1}-\H\l(x)^{\mu+1}] + \z\int
  dx\,\H\l(x)^{\mu+1}}\ .
\label{minima_density}
\ee

We expect as in the hard particle case that iteration of the above
recursion over $l$ either gives 
$l$-independent values (disordered phase) or alternating layers
(ordered phase): $\H^{(2k)}(x)\to \even{\H}(x)$, $\H^{(2k+1)}(x)\to
\odd{\H}(x)$, and similarly for $\I\l(x)$, $\lambda\l$ and $\n\l$. It
is convenient to study directly the ordered case since it includes the
other. The fixed point equations that one needs to solve are then
\bea
\lodd^\mu\ddx\even{\H}(x) &=& -\odd{\I}(x)^\mu
\\
\lodd^\mu\ddx\even{\I}(x) &=& (\z-1)\odd{\H}(x)^\mu
\\
\leven^\mu\ddx\odd{\H}(x)   &=& -\even{\I}(x)^\mu
\\
\leven^\mu\ddx\odd{\I}(x)   &=& (\z-1)\even{\H}(x)^\mu
\eea
with the boundary conditions\eq{boundary_H_I} holding for the
functions in both the even and odd layers, and with
\be
\leven=\even{\I}(1), \qquad
\lodd=\odd{\I}(1)\ .
\ee
A very useful property that follows by combining the fixed point
conditions is 
\be
(\z-1)\lodd^\mu\even{\H}(x)^{\mu+1}+
\leven^\mu\odd{\I}(x)^{\mu+1}=\leven^\mu\lodd^{\mu+1}
\label{conservation}
\ee
independently of $x$: the $x$-derivative of the l.h.s.\ vanishes, and
the value on the r.h.s.\ can be obtained by setting $x=1$. This
identity allows one to decouple the fixed point conditions for
$\even{\H}$ and $\odd{\I}$, giving for the former
\bea
\ddx\even{\H} &=&
-\left[1-(\z-1)\leven^{-\mu}\lodd^{-1}
\even{\H}^{\mu+1}\right]^{\mu/(\mu+1)}\ .
\label{dHdx}
\eea
Integrating by separation of variables and using the boundary
condition $\even{\H}(1)=0$ yields the following implicit expression for
$\even{\H}(x)$
\be
\fl (\mu+1)(1-x)
\left(\frac{\z-1}{\leven^\mu\lodd}\right)^{1/(\mu+1)}
=
B\left(\frac{1}{\mu+1},\frac{1}{\mu+1};(\z-1)\leven^{-\mu}
\lodd^{-1}\even{\H}(x)^{\mu+1}\right)
\label{Heven_cond}
\ee
with $B(p,q;a)=\int_0^a dt\,t^{p-1}(1-t)^{q-1}$ the incomplete Beta
function. For $\odd{\H}(x)$ one has the analogous result with $\leven$
and $\lodd$ swapped.

It now remains to find $\leven$ and $\lodd$. To this end one can
exploit the remaining conditions $\even{\H}(0)=\even{\I}(0)$,
$\odd{\H}(0)=\odd{\I}(0)$, 
from\eq{boundary_H_I}. Combining with\eq{conservation} at $x=0$ and
the corresponding relation with even and odd layers swapped, we find
\be
\even{\H}(0)^{\mu+1}=\even{\I}(0)^{\mu+1}=\leven^{\mu+1}
\frac{(\z-1)(\lodd/\leven)-1}{\z(\z-2)}
\ee
and similarly for the odd layers. Inserting back into\eq{Heven_cond}
for $x=0$ gives
\be
\fl(\mu+1)
\left(\frac{\z-1}{\leven^\mu\lodd}\right)^{1/(\mu+1)}
=
B\left(\frac{1}{\mu+1},\frac{1}{\mu+1};(\z-1)
\frac{\z-1-\leven/\lodd}{\z(\z-2)}
\right)
\label{lambda_even_odd_relation}
\ee
%
%
The same relation again also holds with even and odd layers
swapped. Together, these two conditions determine $\lodd$ and
$\leven$. In the disordered phase, where
$\lodd=\leven$, $\leven$ can be trivially found
from\eq{lambda_even_odd_relation} in closed form (and is equal to
$\Lambda(0)$ as defined below).

For the densities, equation\eq{minima_density} suggests that one
might need the explicit forms of
$\even{\H}(x)$, $\even{\I}(x)$, $\odd{\H}(x)$ and
$\odd{\I}(x)$. However, after some algebra one gets, by transforming
integrals over $x$ to integrals over $\even{\H}$ using\eq{dHdx} and
similarly for integrals involving $\even{\I}$, expressions that depend
only on $\leven$ and 
$\lodd$, and indeed only on their log-ratio $r=\ln(\leven/\lodd)$:
\be
\fl\even{\n} = \frac{\frac{\z}{\z-1}
B\left(\frac{\mu+2}{\mu+1},\frac{1}{\mu+1};
(\z\!-\!1)\frac{\z-1-e^{-r}}{\z(\z-2)}\right)}
{B\left(\frac{\mu+2}{\mu+1},\frac{1}{\mu+1};
(\z\!-\!1)\frac{\z-1-e^{-r}}{\z(\z-2)}\right)
+e^{-2r/(\mu+1)} B\left(\frac{1}{\mu+1},\frac{\mu+2}{\mu+1};
(\z\!-\!1)\frac{\z-1-e^r}{\z(\z-2)}\right)}
\label{min_density_even}
\ee
with an analogous expression for $\odd{\n}$. In the disordered phase,
where $r=0$, the single density is then given by the relatively simple
equation of state
\be
\even{\n} = \frac{\z}{\z-1}\ 
\frac{B\left(\frac{\mu+2}{\mu+1},\frac{1}{\mu+1};\frac{\z-1}{\z}\right)}
{B\left(\frac{1}{\mu+1},\frac{1}{\mu+1};\frac{\z-1}{\z}\right)}\ .
\ee

To understand the solutions for $\leven$, $\lodd$ in the ordered
phase, it is useful to have
a single condition for $r$. Equation\eq{lambda_even_odd_relation} gives
$\leven=\Lambda(r)$ with
\be
\fl\Lambda(r)=(\mu+1)\left[(\z-1)e^{r}\right]^{1/(\mu+1)}
B^{-1}\left(\frac{1}{\mu+1},\frac{1}{\mu+1};(\z-1)
\frac{\z-1-e^{r}}{\z(\z-2)}\right)\ .
\ee
The swapped relation gives $\lodd=\Lambda(-r)$ or
$\leven=e^{r}\Lambda(-r)$. Since the two expressions for
$\leven$ have to agree, the desired condition on $r$ is
\be
e^{-r/2}\Lambda(r)-e^{r/2}\Lambda(-r)=0
\label{r_cond}\ .
\ee
The disordered phase has $r=0$, which is the trivial solution. The
bifurcation to the ordered phase takes place when the first
$r$-derivative at $r=0$ vanishes, \ie\ when $\left.(2\partial_r
\Lambda - \Lambda)\right|_{r=0}=0$. This gives the condition
\be
B\left(\frac{1}{\mu+1},\frac{1}{\mu+1};\frac{\z\c-1}{\z\c}\right)
= \frac{\mu+1}{\mu-1}\ 
\frac{2\left[(\z\c-1)\z\c^{\mu-1}\right]^{1/(\mu+1)}}{\z\c-2}
\label{min_Bethe_zc}
\ee
for the critical value $\z\c$ of the fugacity. It is easy to see that $r$
initially departs from 0 as $(\z-\z\c)^{1/2}$ as $\z$ is increased
to above $\z\c$; this follows because\eq{r_cond} is odd in $r$ so when
the first derivative vanishes the leading term is third order in
$r$. The densities\eq{min_density_even} then have the same
leading-order square root singularity. At generic finite connectivity
$\mu+1$ ($>2$) we therefore retrieve, as hoped, an ordering phase transition
with a standard mean field order parameter exponent $\beta=1/2$.

For generic $\z$ and $\mu$ one needs to solve numerically for $r$
from\eq{r_cond} and then calculate the densities $\even{\n}$ and
$\odd{\n}$ from\eq{min_density_even} and its analogue with even and
odd layers swapped, \ie\ $r\to -r$. Further analytical progress can
again be made for large $\mu$, however. We will need in particular the
scaling of $\z\c$ for large $\mu$. One uses that
\be
B\left(\frac{1}{\mu+1},\frac{1}{\mu+1};a\right)-(\mu+1)\to
\ln[a/(1-a)]
\label{Blimit_fixed_a}
\ee
for $\mu\to\infty$ and $a=\order(1)$ fixed
(see\eq{Beta_order1}). The r.h.s.\
of\eq{min_Bethe_zc} must then also diverge for $\mu\to\infty$, hence
$\z\c\to 2$. To leading order the l.h.s.\ is $\mu+\order(1)$
while the r.h.s.\ is $4/(\z\c-2)$; this forces
\be
\z\c=2+4/\mu+\order(1/\mu^2)
\label{min_zc_large_mu}
\ee
for large $\mu$.

\subsubsection{Large $\mu$, above the transition}

We need to find first how $r$ scales for fixed $\z>\z\c$ and
$\mu\to\infty$. Let us take $r>0$ for definiteness since solutions
come in pairs $(r,-r)$, and write\eq{r_cond} as
\be
e^{r(1-\mu)/(\mu+1)}
= \frac{B\left(\frac{1}{\mu+1},\frac{1}{\mu+1};(\z-1)
\frac{\z-1-e^r}{\z(\z-2)}\right)}
{B\left(\frac{1}{\mu+1},\frac{1}{\mu+1};(\z-1)
\frac{\z-1-e^{-r}}{\z(\z-2)}\right)}
\label{r_cond2}
\ee
For large $\mu$, the l.h.s.\ becomes $e^{-r}$; for this to be $<1$,
the third arguments of the Beta functions on the right cannot stay
bounded away from 0 or 1 since otherwise their ratio would converge to
unity from\eq{Blimit_fixed_a}. The third argument of the numerator
Beta function thus has to approach zero, \ie\ $e^r=z-1-\delta r$ with
$\delta r\to 0$. To leading order we have then
\bea
\frac{1}{\z-1} &=&
\frac{B\left(\frac{1}{\mu+1},\frac{1}{\mu+1};\frac{(\z-1)
\delta r}{\z(\z-2)}\right)}
{B\left(\frac{1}{\mu+1},\frac{1}{\mu+1};1-\frac{\delta
r}{(\z-1)\z(\z-2)}\right)}
\\
&=&\frac{B\left(\frac{1}{\mu+1},\frac{1}{\mu+1};\frac{(\z-1)\delta r}
{\z(\z-2)}\right)}
{B\left(\frac{1}{\mu+1},\frac{1}{\mu+1}\right) -
B\left(\frac{1}{\mu+1},\frac{1}{\mu+1};\frac{\delta
    r}{(\z-1)\z(\z-2)}\right)}\ .
\eea
Now for $a$ remaining finite or going to zero,
$(\mu+1)^{-1}B\left(\frac{1}{\mu+1},\frac{1}{\mu+1};a\right)\to
a^{1/(\mu+1)}$ for $\mu\to\infty$ (see\eq{Beta_exponential_a}).
Bearing in mind that the complete Beta function obeys
$(\mu+1)^{-1}B\left(\frac{1}{\mu+1},\frac{1}{\mu+1}\right) \to 2$, we
get to leading order
\be
\frac{1}{\z-1} = \frac{\delta r^{1/(\mu+1)}}{2-\delta r^{1/(\mu+1)}}
\ee
or, calling $\alpha$ the limiting value of $\delta
r^{1/(\mu+1)}$, $\alpha=2/\z$; $\delta r$ thus decays exponentially
with $\mu$ as $\delta r \sim (2/\z)^\mu$.

The density in the even layers can now be worked out
from\eq{min_density_even}. To leading order, using that
$B\left(\frac{\mu+2}{\mu+1},\frac{1}{\mu+1}; 1-a\right) =
B\left(\frac{\mu+2}{\mu+1},\frac{1}{\mu+1}\right) -
B\left(\frac{1}{\mu+1},\frac{\mu+2}{\mu+1};a\right)\to
(\mu+1)(1-a^{1/(\mu+1)})$ (see after\eq{Beta_exponential_a})
\bea
\even{\n} &=& \frac{\frac{\z}{\z-1}
B\left(\frac{\mu+2}{\mu+1},\frac{1}{\mu+1};
1-\frac{\delta r}{(\z-1)\z(\z-2)}\right)}
{B\left(\frac{\mu+2}{\mu+1},\frac{1}{\mu+1};
1-\frac{\delta r}{(\z-1)\z(\z-2)}\right)
+B\left(\frac{1}{\mu+1},\frac{\mu+2}{\mu+1};
\frac{(\z-1)\delta r}{\z(\z-2)}\right)}
\\
&=& \frac{\z}{\z-1}\ \frac{1 - \alpha}{
1-\alpha + \alpha} \ = \ \frac{\z-2}{\z-1}\ .
\eea
This agrees with the simple two-box mean field theory as we had hoped.

The density in the odd layers, on the other hand, goes to zero for
large $\mu$:
\bea
\odd{\n} &=& \frac{\frac{\z}{\z-1}
B\left(\frac{\mu+2}{\mu+1},\frac{1}{\mu+1};
\frac{(\z-1)\delta r}{\z(\z-2)}\right)}
{B\left(\frac{\mu+2}{\mu+1},\frac{1}{\mu+1};
\frac{(\z-1)\delta r}{\z(\z-2)}\right)
+B\left(\frac{1}{\mu+1},\frac{\mu+2}{\mu+1};
1-\frac{\delta r}{(\z-1)\z(\z-2)}\right)}
\\
&\approx& \frac{\delta r}{(\z-2)(\mu+1)}\ .
\eea
This is exponentially small in $\mu$ because $\delta r$ is.

We discuss briefly the behaviour for large $\mu$ of $\Ss_0(x)$ and
$\Ss_1(x)$. Up to an overall normalization factor these give the
probabilities, in the random landscape ensemble weighted by the
fugacity $z$, that the root node of a Cayley tree has random function 
value $x$ and is (for $\Ss_1(x)$) or is not (for $\Ss_0(x)$) a minimum. For
large $\mu$, the fact that the root node has $\mu$ rather than $\mu+1$
neighbours becomes unimportant and these probabilities also apply to
an arbitrary node in the bulk of the Bethe lattice.

We find by constructing the explicit solutions for $\H(x)$ etc.\ that
there is a threshold value of $x$, $x^*=(\z-2)/[2(\z-1)]$ so that for
$\mu\to\infty$ one has $\even{\Ss}_0(x)=\odd{\Ss}_0(x)=\Theta(x-x^*)$,
$\even{\Ss}_1(x)=\z\Theta(x^*-x)$, $\odd{\Ss}_1(x)=0$. (Correspondingly,
the functions $\H$ and $\I$ are piecewise linear below and above
$x^*$.) So in the even layers, \ie\ those with a nonzero density of
minima, a site is a minimum if its value $x_i$ is below $x^*$, and not
a minimum otherwise. In the odd layers, no sites are minima, and the
values $x_i$ at all sites are above $x^*$. This is exactly the same
phenomenology as in the two-box mean field theory, confirming again
that the latter captures most of the physics of the large connectivity
limit on the Bethe lattice. The exception is the region around the
ordering transition where the square-root singularities and hence the
order parameter exponent $\beta=1/2$ are visible: this becomes
vanishingly small as we will now see.

\subsubsection{Large $\mu$, around the transition}

From the large-$\mu$ expansion\eq{min_zc_large_mu} we expect that the
appropriate scaling for the fugacity in the region around the phase
transition is $\z=2+\tilde\z/\mu$. As we will see, this corresponds to
$r$ being of order $1/\mu$, $r=\tilde r/\mu$. With these scalings, the
third arguments of the Beta functions in\eq{r_cond2} become
$\half(1\mp\tilde r/\tilde\z)$. Using from\eq{Beta_order1} that for
$\mu\to\infty$,
$B\left(\frac{1}{\mu+1},\frac{1}{\mu+1};\half(1+a)\right)-(\mu+1) \to
\ln[(1+a)/(1-a)]$ and keeping only terms of $\order(1)$ and
$\order(1/\mu)$ gives
\be
\fl 1-\frac{\tilde r}{\mu}
= \frac{\mu+1+\ln[(1-\tilde r/\tilde\z)/(1+\tilde r/\tilde\z)]}
{\mu+1+\ln[(1+\tilde r/\tilde\z)/(1-\tilde r/\tilde\z)]}
= 1+\frac{2}{\mu}\ln[(1-\tilde r/\tilde\z)/(1+\tilde r/\tilde\z)]\ .
\ee
Equating the $\order(1/\mu)$ terms shows
\be
\tilde r = 
2\ln[(1+\tilde r/\tilde\z)/(1-\tilde r/\tilde\z)] = 4\,\mbox{artanh}(\tilde
r/\tilde\z)
\ee
or 
\be
\tilde\z = \frac{\tilde r}{\tanh(\tilde r/4)}\ .
\label{ztilde_vs_rtilde}
\ee
The critical point is reached for $\tilde r\to 0$, giving
$\tilde\z\c=4$ in agreement with\eq{min_zc_large_mu}.

\begin{figure}
\begin{center}
\includegraphics[width=10cm]{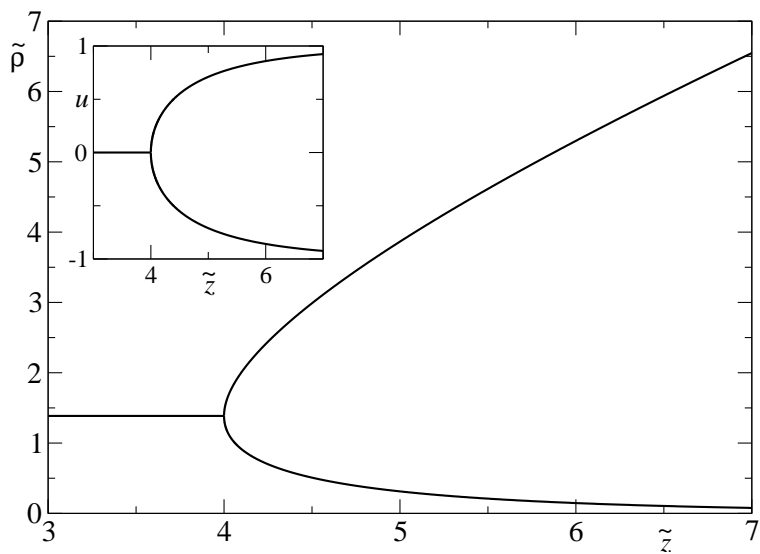}
\end{center}
\caption{Equation of state of the random minima problem around the
  ordering transition, for the Bethe lattice in the limit of large
  connectivity $\mu+1$. Shown are the scaled densities of minima,
  $\tilde\n$, in the two boxes, against the scaled fugacity $\tilde
  z$. Inset: After transforming nonlinearly to $u=1-2e^{-\tilde\n/2}$,
  the equation of state becomes that of a mean field ferromagnet at
  inverse temperature $\tilde z/4$.
\label{fig:eqn_of_state}
}
\end{figure}

It remains to work out the densities. In the
expression\eq{min_density_even} for the even layers, the arguments of
the Beta functions simplify as before, and also $\z/(\z-1)\to 2$, so
that
\be
\even{\n} = \frac{2
B\left(\frac{\mu+2}{\mu+1},\frac{1}{\mu+1};
\half\left(1+\frac{\tilde r}{\tilde\z}\right)\right)}
{B\left(\frac{\mu+2}{\mu+1},\frac{1}{\mu+1};
\half\left(1+\frac{\tilde r}{\tilde\z}\right)\right)
+
B\left(\frac{1}{\mu+1},\frac{\mu+2}{\mu+1};
\half\left(1-\frac{\tilde r}{\tilde\z}\right)\right)}\ .
\ee
The second Beta function in the numerator equals $\mu+1$ to leading order
while the other ones are $\order(1)$,
$B\left(\frac{\mu+2}{\mu+1},\frac{1}{\mu+1};a\right)\to \int_0^a
dt\,(1-t)^{-1}=-\ln(1-a)$, so that the scaled density
$\tilde\even{\n}=\even{\n}\mu$ becomes for $\mu\to\infty$
\be
\tilde\even{\n} = -2\ln\left[\half\left(1-\frac{\tilde
r}{\tilde\z}\right)\right]
\label{tilde_rho_even}
\ee
and similarly in the odd layers, after swapping $\tilde r\to -\tilde
r$,
\be
\tilde\odd{\n} = -2\ln
\left[\half\left(1+\frac{\tilde r}{\tilde\z}\right)\right]\ .
\label{tilde_rho_odd}
\ee
The
equations~(\ref{ztilde_vs_rtilde},\ref{tilde_rho_even},\ref{tilde_rho_odd})
give the equation of state in the phase transition region. The
occurrence of the auxiliary parameter $\tilde r$ is a little awkward
but can be eliminated if we transform the (scaled) densities
nonlinearly as
\be
\even{u} = 1-2e^{-\tilde\n/2}, \qquad
\odd{u} = 1-2e^{-\tilde\odd{\n}/2}
\label{u_def}
\ee
so that, from~(\ref{tilde_rho_even},\ref{tilde_rho_odd}),
$\even{u}=-\odd{u}$ always.
%
%
By combining~(\ref{ztilde_vs_rtilde},\ref{tilde_rho_even}) one then
sees that
\be
\even{u} = \frac{\tilde r}{\tilde\z}=\tanh(\tilde r/4)
\ee
and so finally $\tilde\z = \tilde
r/\even{u}=4\,\mbox{artanh}(\even{u})/\even{u}$ or
\be
\even{u}=\tanh((\tilde\z/4)\even{u})\ .
\ee
Since $\even{u}=-\odd{u}$, the same equation also holds for the
density in the odd layers. Remarkably, therefore, once the densities
are nonlinearly transformed according to\eq{u_def}, they depend on the
fugacity exactly as the magnetizations in a mean field ferromagnet
with unit interaction strength and inverse temperature $\tilde\z/4$.
We show the equation of state in Fig.~\ref{fig:eqn_of_state}, both in
terms of the (scaled) densities $\tilde\n$ and, in the inset, the
transformed variables $u$.

\subsection{Mean field theory with soft minima}
\label{sec:Bethe_soft}

In this final subsection we ask whether the behaviour around the
ordering transition that we found for a highly connected Bethe lattice
can also be obtained directly within a mean field theory. It turns out
that this is possible: drawing inspiration from our treatment of
the hard particle model, we make the labelling of sites as minima
`soft'.

The two-box setup is initially the same as for hard particles, with
generating function
\be
\G(z,N) = \lav \z^{\Me+\Mo} \rav =
\lav \z^{\sum_i(\even{\m}_i+\odd{\m}_i)} \rav
\ee
Here the average is over our random landscape ensemble as before,
while $\Me$ and $\Mo$ label the total number of minima in the left and
right box, respectively. For hard minima of course one and only one of
these quantities is ever nonzero; for soft minima both $\Me$ and $\Mo$
can be nonzero.

To define `soft' minima, we first introduce an auxiliary variable
$\sig_i\in\{0,1\}$ at each site. We can obtain the usual generating
function for hard minima by forcing this to be 0 if $\m_i=0$; otherwise
we allow it to be 0 or 1. Assigning weight factors 1 and $\z-1$ to
$\sig_i=0$ and $1$, respectively, we can then write the factor from
each site in the generating function as
\be
z^{\m_i} = \sum_{\sig_i=0,1} (z-1)^{\sig_i} \delta_{\sig_i(1-\m_i),0}\ .
\ee
(Indeed, for $\m_i=0$ only $\sig_i=0$ is allowed and we get
$z^0=1=(z-1)^0$; in the opposite case we have $z^1=(z-1)^0+(z-1)^1$.)
Now to make the minima soft, we relax the constraint that $\sig_i=0$ if
$\m_i=0$, \ie\ we replace $\delta_{\sig_i(1-\m_i),0} \to
\exp[-\alpha\sig_i(1-\hat\m_i)]$. Here
\be
\hat\m_i=N^{-1}\sum_{j=1}^\N \Theta(\odd{x}_j-\even{x}_i)
\label{soft_min_ind}
\ee
is a soft version of $\m_i$: it measures what fraction of numbers in
the other box are above $\even{x}_i$, so that
$\hat\m_i=0,1/N,\ldots,1-1/N$ corresponds to $\m_i=0$ and $\hat\m_i=1$
to $\m_i=1$. Thus, when $\m_i=0$ we have $1-\hat\m_i\geq 1/N$ and for
$\alpha\to\infty$ at fixed $N$ our soft minima weight
$\exp[-\alpha\sig_i(1-\hat\m_i)]$ reverts to
$\delta_{\sig_i(1-\m_i),0}$ as it should. As in the hard sphere case
we in fact take $N\to\infty$ first and then $\alpha\to\infty$.

We summarize our starting point: the generating function for soft
minima is
\bea
\G(z,N) &=& \Tr (\z-1)^{\sum_{i=1}^\N (\even{\sig}_i + \odd{\sig}_i)} e^{\N
A}
\label{min_soft_partition_fn}
\\
A &=& 
\frac{1}{\N}\ln 
\lav \exp\left( -\alpha\sum_i[\even{\sig}_i
(1-\hat\even{\m}_i) + \odd{\sig}_i (1-\hat\odd{\m}_i)]\right) \rav
\label{A_def}
\eea
where $\Tr$ abbreviates the sum over all $\even{\sig}_i$ and
$\odd{\sig}_i$. The soft version of the minimum
indicator variables $\odd{\m}_i$ in the right box is defined in the
obvious way by swapping the roles of $\even{x}$ and $\odd{x}$ in
(\ref{soft_min_ind}), i.e.\
$\hat\odd{\m}_i=N^{-1}\sum_{j=1}^\N \Theta(\even{x}_j-\odd{x}_i)$.

To calculate $\G$, consider first the average
from\eq{A_def}. By permutation symmetry within each
box, this can only depend on the numbers $\even{\T}$, $\odd{\T}$ of
nonzero $\sig$'s in the two boxes. Writing out the definition of the
$\hat\even{\m}_i$ and $\hat\odd{\m}_i$ in terms of sign functions, this gives
\bea
\fl A
&=& \frac{1}{\N}\ln\lav \exp\left(-\frac{\alpha}{2\N}
\left[\sum_{i=1}^{\even{\T}}\sum_{j=1}^\N
(1+\sgn(\even{x}_i-\odd{x}_j))
+ \sum_{j=1}^{\odd{\T}}\sum_{i=1}^\N
(1-\sgn(\even{x}_i-\odd{x}_j))\right]
\right) \rav
\\
\fl &=& -\alpha \frac{\even{\T}+\odd{\T}}{2\N}+
b(\even{\T},\N-\odd{\T}) + b(\N-\even{\T},\odd{\T})
\label{A_in_terms_of_b}
\eea
Here we have used that the sgn terms with $1\leq j\leq \odd{\T}$ in the
first sum exactly cancel those with $1\leq i\leq \even{\T}$ in the
second one, so that the remaining average
factorizes into two independent terms of the form
\be
e^{\N b(\Ta,\Tb)} = \lav \exp\left(-\frac{\alpha}{2\N}
\sum_{i=1}^{\Ta} \sum_{j=1}^{\Tb} \sgn(\even{x}_i-\odd{x}_j)\right)
\rav\ .
\label{b_def}
\ee
The replacement $\even{x}_i\to 1-\even{x}_i$, $\odd{x}_j\to
1-\odd{x}_j$ leaves the distribution of these variables unchanged,
hence $b$ is symmetric under $\alpha\to -\alpha$, as well as under
interchange of $\Ta$ and $\Tb$. To evaluate $b$, we can assume without
loss of generality that the $\even{x}_i$ are ordered. Using also that
the average over the $\odd{x}_j$ factorizes,
\bea
\fl e^{\N b(\Ta,\Tb)} &=& \biggl\langle \left[
\even{x}_1 e^{-\Ta \alpha/2\N} + (\even{x}_2-\even{x}_1) e^{(2-\Ta)
\alpha/2\N} + \ldots + (\even{x}_{\Ta}-\even{x}_{\Ta-1})
e^{(\Ta-2i)\alpha/2\N} \right.
\nonumber\\
\fl & & \left. {}+{} (1-\even{x}_{\Ta})
e^{\Ta\alpha/2\N}\right]^{\Tb}\biggr\rangle
\eea
where the remaining average is over the $\even{x}_i$. Denote the
quantity raised to the power $\Tb$ by $y$. Setting also $\u_0=\even{x}_1$,
$\u_1=\even{x}_2-\even{x}_1$, \ldots,
$\u_{\Ta-1}=\even{x}_{\Ta}-\even{x}_{\Ta-1}$, $\u_{\Ta}=1-\even{x}_{\Ta}$, 
the $\u_i$ are non-negative (because of the ordering of the
$\even{x}_i$) and uniformly distributed apart from the constraint
$\sum_{i=0}^{\Ta} \u_i=1$, such that $P(\{\u_i\}) =
\Ta!\,\delta(1-\sum_i \u_i)$. The characteristic function of $y$ is then
\bea
\fl \lav e^{\N\omega y} \rav &=& \Ta! \int \frac{\N d\lambda}{2\pi i}
\int_0^\infty \prod_{i=0}^{\Ta} d\u_i\, \exp\left(
\N\lambda(1-\sum_i \u_i) +\N\omega \sum_i \u_i e^{(2i-\Ta)\alpha/2\N}\right)
\\
\fl &=&\Ta! \int \frac{\N d\lambda}{2\pi i} e^{\N\lambda}
\prod_{i=0}^{\Ta} \left(\N\lambda - \N\omega
  e^{(i-\Ta/2)\alpha/\N}\right)^{-1}\ .
\eea
Reverse Fourier transforming now produces
\bea
\fl e^{\N b(\Ta,\Tb)} = \lav y^{\Tb} \rav 
&=& \Ta! \int dy \int \frac{\N d\omega}{2\pi i}
\int \frac{\N d\lambda}{2\pi i} 
\exp\biggl[\Tb\ln y - \N\omega y + \N\lambda
\nonumber\\
\fl & & {}-{}(\Ta+1)\ln\N - \sum_{i=0}^{\Ta} \ln\left(\lambda - \omega
e^{(i-\Ta/2)\alpha/\N}\right)\biggr]\ .
\eea
Defining the intensive quantitites $\nsa=\Ta/\N$, $\nsb=\Tb/\N$, we
can do the integral using steepest descents for $\N\to\infty$:
\be
\fl b(\nsa,\nsb) = \max_{y,\omega,\lambda} \left\{\nsa\ln(\nsa/e) + \nsb\ln y -
\omega y + \lambda - \int_{-\nsa/2}^{\nsa/2} d\w\,\ln(\lambda - \omega
e^{\alpha\w})\right\}\ .
\label{b_SP}
\ee
Setting the derivatives w.r.t. $y$, $\omega$ and $\lambda$ to zero
gives the saddle point equations
\be
\omega = \frac{\nsb}{y}, \qquad
y = \int_{-\nsa/2}^{\nsa/2} d\w\,\frac{e^{\alpha\w}}{\lambda - \omega
e^{\alpha\w}}, \qquad
1 = \int_{-\nsa/2}^{\nsa/2} d\w\,\frac{1}{\lambda - \omega
e^{\alpha\w}} \ .
\ee
Combining the last two we find
\be
y = \frac{1}{\omega}\int_{-\nsa/2}^{\nsa/2} d\w\,\left(\frac{\lambda}
{\lambda - \omega e^{\alpha\w}}-1\right) = \frac{\lambda-\nsa}{\omega}
\ee
and hence $\lambda = \nsa+\nsb$. In the last saddle point equation
we can perform the integral explicitly, yielding
\be
1 = \frac{1}{\alpha\lambda}\ln\left(
\frac{\lambda e^{\alpha\nsa/2}-\omega}{\lambda e^{-\alpha\nsa/2}-\omega}
\right)
\ee
and we can solve for $\omega$:
\be
\omega = (\nsa+\nsb)
\frac{e^{\alpha\nsb/2}-e^{-\alpha\nsb/2}}{e^{\alpha(\nsa+\nsb)/2} -
e^{-\alpha(\nsa+\nsb)/2}}
\ee
Together with $\lambda=\nsa+\nsb$, $y=\nsb/\omega$ we thus have all
saddle point values explicitly. The derivatives
of $b(\nsa,\nsb)$ that we will need become
\bea
\frac{\partial b}{\partial \nsa} &=& \ln\nsa - \half \ln\left[
(\lambda-\omega e^{\alpha \nsa/2})(\lambda-\omega e^{-\alpha
\nsa/2})\right]
\\
&=& \ln\frac{\nsa}{\nsa+\nsb} + \frac{\alpha\nsb}{2} + \ln\left(
\frac{1-e^{-\alpha(\nsa+\nsb)}}{1-e^{-\alpha\nsa}}\right)
\label{b_der}
\eea
and $\partial b/\partial \nsb$ has the same form with $\nsa$ and
$\nsb$ interchanged. This symmetry property is clear from the
definition\eq{b_def} but not so obvious from the saddle point
representation\eq{b_SP}. An explicitly symmetric expression can be
obtained from\eq{b_der} by integrating from $t_1=0$ (where $b=0$);
after a little algebra, this can be cast in the form
\be
\alpha b(\nsa,\nsb) = F(\alpha(\nsa+\nsb))-F(\alpha\nsa)-F(\alpha\nsb)
\ee
with $F(x)=\int_0^x du\,\ln[2\sinh(u/2)/u]$. One might hope that a
derivation exists which directly reveals this simple structure, but so
far we have been unable to find one.

Now we can finally write down the saddle point equations for the full
generating function. In terms of the $\sig$-densities $\nse=\even{\T}/\N$,
$\nso=\odd{\T}/\N$, one has from
(\ref{min_soft_partition_fn},\ref{A_def},\ref{A_in_terms_of_b})
\bea
\fl\frac{1}{\N} \ln \G &=& \max_{\nse,\nso} \left\{
\entropy{\nse} + \entropy{\nso} +
(\nse+\nso)\ln(\z-1)+ A\right\} \\
\fl &=& \max_{\nse,\nso} \left\{
\entropy{\nse} + \entropy{\nso} +
(\nse+\nso)[\ln(\z-1)-\alpha/2] + b(\nse,1-\nso) + b(1-\nse,\nso)
\right\}
\eea
where $\entropy{\nse}$ and $\entropy{\nso}$ again account for the combinatorial
(entropic) contributions. Then
\bea
\fl \frac{\partial}{\partial\nse} \frac{1}{\N} \ln \G &=&
\ln\frac{1-\nse}{\nse} + \ln(\z-1)-\alpha/2
+ \left. \frac{\partial b}{\partial \nsa}\right|_{\nsa=\nse,\,\nsb=1-\nso} 
- \left. \frac{\partial b}{\partial \nsa}\right|_{\nsa=1-\nse,\,\nsb=\nso}
\\
\fl &=& 
\ln(\z-1)-\alpha\nso+\ln\left(\frac{1-\nse+\nso}{1+\nse-\nso}\right)
+\ln\left(\frac{1-e^{-\alpha(1+\nse-\nso)}}{1-e^{-\alpha(1-\nse+\nso)}}
\ \frac{1-e^{-\alpha(1-\nse)}}{1-e^{-\alpha\nse}}\right)
\eea
and this must vanish at the saddle point. The corresponding equation for
$\nso$ just has $\nse$ and $\nso$ swapped.

In the disordered phase $\nse=\nso$, the saddle point equation can be solved
explicitly to get $\ns=\alpha^{-1}\ln[\z/(1+e^{-\alpha})]$.  One can
then again ask about bifurcations to solutions where $\nse\neq
\nso$. The critical value of $\z$ can be got as follows: think of the
first saddle point equation as defining implicitly $\nso$ as a function of
$\nse$; the second saddle point equation defines the inverse function, which
graphically is flipped about the diagonal. The disordered fixed point
on the diagonal becomes unstable when the slope $d\nso/d\nse=-1$. The
resulting condition on $\z\c$ looks complicated, but neglecting terms
that are exponentially small in $\alpha$ one gets $\z\c =
2(1-\alpha^{-1})^2/(1-4\alpha^{-1}) =
2+4\alpha^{-1}+\order(\alpha^{-2})$. This looks encouraging: with the
identification $\mu\equiv\alpha$, it is identical to the
result\eq{min_zc_large_mu}, suggesting that the soft minima mean field
theory captures the large connectivity limit on the Bethe lattice.

We recall for the evaluation in the following subsections that
$\nse=\N^{-1}\sum_i\even{\sig}_i$, and similarly $\nso$, are the
densities of the $\sig$-variables. For large enough $\alpha$, we can
have $\sig_i=1$ only when there is genuinely a minimum at site $i$
($\m_i=1$); but even if $\m_i=1$ then $\sig_i=0$ with probability
$1/(1+\z-1)=1/\z$. So for $\alpha\to\infty$ the $\sig$-densities are
related to the true densities of minima by
$\nse=[(\z-1)/\z]\even{\n}$, $\nso=[(\z-1)/\z]\odd{\n}$.

\subsubsection{Large $\alpha$, above the transition}

Here we expect that one of the two densities (say $\nse$) stays finite
and $<1$ while the other ($\nso$) goes to zero. The saddle point 
equations are then, up to exponentially small terms:
\bea
0&=&\ln(\z-1)-\alpha\nso+\ln\left(\frac{1-\nse+\nso}{1+\nse-\nso}\right)
\\
0&=&\ln(\z-1)-\alpha\nse-\ln\left(\frac{1-\nse+\nso}{1+\nse-\nso}\right)
-\ln(1-e^{-\alpha\nso})\ .
\eea
In the second equation, the only way to balance the $-\alpha\nse$ term
is to have $\nso$ vanish faster than $1/\alpha$ so that the argument
of the last log tends to zero; the log itself can then be approximated
as $-\ln(\alpha\nso)$. This gives to leading order $\nso\sim
\alpha^{-1}e^{-\alpha\nse}$. Inserting into the first equation then
leads to $(1+\nse)/(1-\nse)=\z-1$ or $\nse=(\z-2)/\z$. The density of
minima in this box is therefore
$\even{\n}=[\z/(\z-1)]\nse=(\z-2)/(\z-1)$, consistent with our direct
calculation in the hard minima limit. The critical fugacity is
$\z\c=2$, also as expected. Overall the $\alpha\to\infty$
limit correctly reproduces the hard minima scenario as desired.

\subsubsection{Large $\alpha$, around the transition}

By analogy with the Bethe lattice calculation, we scale the
$\sig$-densities and the fugacity as $\nse=\tilde\nse/\alpha$,
$\nso=\tilde\nso/\alpha$ and $\z=2+\tilde\z/\alpha$, respectively. The
saddle point equations are then, again up to exponentially small terms,
\bea
0&=&\ln(\z-1)-\tilde\nso+\ln\left(\frac{1-(\tilde\nse-\tilde\nso)/\alpha}
{1+(\tilde\nse-\tilde\nso)/\alpha}\right) -
\ln\left(1-e^{-\tilde\nse}\right)
\\
0&=&\ln(\z-1)-\tilde\nse-\ln\left(\frac{1-(\tilde\nse-\tilde\nso)/\alpha}
{1+(\tilde\nse-\tilde\nso)/\alpha}\right) -
\ln\left(1-e^{-\tilde\nso}\right)
\eea
It is again useful to make a nonlinear transformation from the
$\sig$-densities to
\be
\even{u}=1-2e^{-\tilde\nse}, \qquad
\odd{u} =1-2e^{-\tilde\nso}\ .
\label{u_def_boxes}
\ee
The first saddle point equation then implicitly defines a function $U$ via
$\odd{u}=U(\even{u})$. The second one gives $\even{u}=U(\odd{u})$. For
these to be consistent with each other, we require
\be
\alpha[U(\even{u})-U^{-1}(\even{u})]=0\ .
\label{u_relation}
\ee
Here $U^{-1}$ is the inverse function of $U$; the factor $\alpha$ will
be useful shortly. To find the function $U$ for large $\alpha$,
we expand the first saddle point equation, keeping terms of
$\order(1)$ and $\order(1/\alpha)$:
\be
\fl 0 = \frac{\tilde\z}{\alpha} - \tilde\nso -
\frac{2(\tilde\nse-\tilde\nso)}{\alpha} -
\ln\left(1-e^{-\tilde\nse}\right)
= \ln\left(\frac{1-\odd{u}}{1+\even{u}}\right) +
\frac{\tilde\z+2\ln[(1-\even{u})/(1-\odd{u})]}{\alpha}\ .
\label{SP_1st}
\ee
To leading order this gives $\odd{u}=U(\even{u})=-\even{u}$: the
function $U$ is identical to its inverse. This is why we need to go to
$\order(1/\alpha)$ to get a nontrivial condition for $\even{u}$, as
emphasized by the factor $\alpha$ in\eq{u_relation}. Now insert the
leading order relation
$\odd{u}=-\even{u}$ into the $\order(1/\alpha)$ term above to get
\bea
\fl \odd{u}=U(\even{u})&=&1-(1+\even{u})\left(1-
\frac{\tilde\z+2\ln[(1-\even{u})/(1+\even{u})]}{\alpha}\right)\\
\fl &=& -\even{u} + (1+\even{u})
\frac{\tilde\z+2\ln[(1-\even{u})/(1+\even{u})]}{\alpha}\ .
\eea
The inverse function is obtained by solving\eq{SP_1st} for $\even{u}$:
\bea
\fl \even{u}=U^{-1}(\odd{u})&=&
-1+(1-\odd{u})\left(1+
\frac{\tilde\z+2\ln[(1+\odd{u})/(1-\odd{u})]}{\alpha}\right)\\
\fl &=& -\odd{u} + (1-\odd{u})
\frac{\tilde\z+2\ln[(1+\odd{u})/(1-\odd{u})]}{\alpha}\ .
\eea
So the equation\eq{u_relation} determining $\even{u}$ becomes for
$\alpha\to\infty$ 
\bea
\fl 0&=&(1+\even{u})
(\tilde\z+2\ln[(1-\even{u})/(1+\even{u})])
-(1-\even{u})
(\tilde\z+2\ln[(1+\even{u})/(1-\even{u})])
\\
\fl &=& 2\tilde\z\even{u} + 4\ln[(1-\even{u})/(1+\even{u})] =
2\tilde\z\even{u} - 8\,\mbox{artanh}(\even{u})
\eea
or
\be
\even{u}=\tanh((\tilde\z/4)\even{u})\ .
\ee
This is exactly as on the Bethe lattice around the transition, so the
entire scaling behaviour in this region matches between the two cases,
namely, the Bethe lattice in the limit of large connectivity $\mu$ and
the soft minima problem in the nearly hard limit of large $\alpha$.
Notice that, while the definitions of the relevant nonlinear
transformations\eq{u_def} and\eq{u_def_boxes} of the density variables
look different, they are in fact identical because
$\tilde\even{\n}=[\z/(\z-1)]\tilde\nse = 2\tilde\nse +
\order(1/\alpha)$.


\section{Summary and outlook}
\label{sec:summary}

In summary, we have analysed the number and distribution of minima in
random landscapes defined on non-Euclidean lattices. Using an ensemble
where random landscapes are reweighted by a fugacity factor $z^M$
depending on the number of minima $M$, the simplest viable (two-box) mean
field theory showed an ordering phase transition at $z\c=2$. For
$z>z\c$, one box contains an extensive number of minima with
density $\even{\n}=(z-2)/(z-1)$. The onset of order seemed to be
governed by an unusual order parameter exponent $\beta=1$, which
motivated our study on the Bethe lattice.

Using recursion techniques, we found a full solution of the problem on
the Bethe lattice which showed that for any finite connectivity
$\mu+1$ ($>2$) there is indeed an ordering transition with a
conventional mean field order parameter exponent $\beta=1/2$. As $\mu$
becomes large, the region around the transition where this behaviour
is visible shrinks as $1/\mu$. It disappears as $\mu\to\infty$ at
fixed fugacity $z$, and this is what causes the unusual effective
exponent in the two-box mean field theory. We analysed separately the
scaling for large $\mu$ for fixed $z$ above the transition and for $z$
within $1/\mu$ of $z\c$. In the latter case, a nonlinear
transformation turns out to map the equation of state neatly onto that
of a mean field ferromagnet. Finally, we showed that the region around
the phase transition can also be analysed directly within a mean field
approach, by making the assignment of minima `soft' and then taking
the nearly hard limit ($\alpha\to\infty$). This was motivated by our
analogous treatment of the hard sphere lattice gas, where a softening
of the nearest neighbour exclusion revealed the ordering phase
transition that remains entirely hidden within the two-box
mean field theory.

In the mean field approach we also considered the joint distribution
of minima and maxima of random landscapes. Here two fugacities enter,
$z_1$ and $z_2$, and in addition to the phase transitions at $z_1=2$
and $z_2=2$ where the number of minima and maxima respectively first
becomes extensive, there is a first-order transition on
the line $1/z_1+1/z_2=1$. Beyond this line, essentially all points in the
landscape are either minima and maxima; in our mean field setup, these
sites are separated into the two boxes.

In future work, it should be possible to extend the analysis of joint
distributions of minima and maxima to the Bethe lattice. This would
presumably require three generating functions, for sites that are
minima, maxima or neither. Generalizing the soft minima/maxima
approach looks less easy because for `soft' labels one would also have
to consider sites that are labelled as both minima and maxima. 
It would also be interesting to generalize further, and consider not
just minima but also nodes with fixed number $k=1, 2, \ldots$ of
lower-lying neighbours.

Finally, one would like to extend our calculation also to large random
graphs with the same local structure as a Bethe lattice, i.e.\ regular
graphs where
all nodes have the same number ($\mu+1$) of neighbours. Given that short loops are rare
on such graphs, one might intuitively expect to see the same
phenomenology. However, the strict sublattice ordering on the Bethe
lattice cannot be maintained in the inevitable presence of at least
some loops with an odd
number of links, and so in actual fact it is likely that one would
instead obtain glassy phases as in related hard particle
models~\cite{HanWei05,RivBirMarMez04}. Generalizing our approach to
this scenario appears to be a challenging problem indeed.

\ack We acknowledge gratefully the hospitality of the Newton
Institute, where this collaboration was initiated.

\appendix

\section{Minima and maxima for $z_1>2$ or $z_2>2$}
\label{app:minmax}

We outline two methods for understanding the two-box problem in the
case where we track both minima and maxima. The first one starts from
the large-deviation form of\eq{PM1M2}. It is easy to see that the
first and third terms can never be larger than the second. Taking $N$
large at fixed densities $\rho_1=M_1/N$ and $\rho_2=M_2/N$ of the
minima and maxima then gives up an irrelevant constant
\be
N^{-1}\ln P(N\rho_1,N\rho_2,N) =
(2-\rho_1-\rho_2)\entropy{\frac{1-\rho_1}{2-\rho_1-\rho_2}}\ .
\ee
If one rewrites the definition\eq{Gz1z2} of the generating function as an
integral over $\rho_1$ and $\rho_2$, the latter will therefore be dominated by
those values maximizing the function
\be
\gamma(\rho_1,\rho_2)=
(2-\rho_1-\rho_2)\entropy{\frac{1-\rho_1}{2-\rho_1-\rho_2}} 
+\rho_1 \ln z_1 + \rho_2 \ln z_2\ .
\ee
Now take for definiteness $z_1>z_2$, so that any maxima will obey
$\rho_1\geq \rho_2$. In this regime we can set
$1-\rho_1=\kappa(2-\rho_1-\rho_2)$ with $0\leq\kappa\leq 1/2$ and
have at fixed $\kappa$ a {\em linear} variation with $1-\rho_2$:
\bea
\gamma(\rho_1,\rho_2)&=&\ln (z_1 z_2) + (1-\rho_2)s(\kappa)
\\
s(\kappa)&=&  - \frac{\kappa}{1-\kappa} \ln(\kappa z_1) -
\ln[(1-\kappa)z_2]\ .
\eea
The slope function $s(\kappa)$ now tells us where the maxima of
$\gamma(\rho_1,\rho_2)$ are. First we maximize over $\kappa$; if the
maximum value of $s(\kappa)$ is positive, we get a maximum of
$\gamma(\rho_1,\rho_2)$ at $\rho_2=0$ and hence
$\rho_1=(1-2\kappa)/(1-\kappa)$, otherwise a maximum at $\rho_2=1$
and $\rho_1=1$.

The derivative of $s(\kappa)$ is $s'(\kappa)=-\ln(\kappa
z_1)/(1-\kappa)^2$. For $z_1<2$ this is always positive and the
maximum is at $\kappa=1/2$, where $s(1/2)=-\ln(z_1 z_2/4)>0$ (given
that $z_2<z_1<2$). So $\gamma(\rho_1,\rho_2)$ is maximal at
$\rho_1=\rho_2=0$, consistent with the analysis in the main text that
showed that in this regime minima and maxima are both intensive in
number.

For $z_1>2$, the maximum of $s(\kappa)$ is at $\kappa=1/z_1$, where
$s(1/z_1)=-\ln[(1-1/z_1)/(1/z_2)]$. If $1/z_2>1-1/z_1$, this value is
positive and $\gamma(\rho_1,\rho_2)$ has a maximum at $\rho_2=0$,
$\rho_1=(1-2\kappa)/(1-\kappa)=(z_1-2)/(z_1-1)$. This is the mixed
regime, with $M_1$ extensive and $M_2$ intensive,
see\eq{minmax_mixed}. For $1/z_2<1-1/z_1$, finally, the maximum value
of $s(\kappa)$ is negative and $\gamma(\rho_1,\rho_2)$ has its maximum
at $\rho_1=\rho_2=1$. This is the fully separated regime, where one
box contains essentially only minima and the other only maxima. Note
that, as stated in the main text, at the first-order transition $1/z_1
+1/z_2=1$, the entire line in the $(\rho_1,\rho_2)$ plane
corresponding to $\kappa=1/z_1$ is degenerate, i.e.\ has the same
value of $\gamma(\rho_1,\rho_2)$. A further peculiarity is 
that there is no metastability: neither of the phases persists as a
local maximum of $\gamma(\rho_1,\rho_2)$ on the corresponding `wrong'
side of the transition line.

It remains to find the average number of maxima in the mixed regime
($z_1>2$, $1/z_1+1/z_2>1$). We already know that $\rho_1$ is nonzero
then; on general grounds its fluctuations ($\sim 1/\sqrt{N}$)
must become negligible for large $N$. We can then take the limit
$N\to\infty$ in\eq{PM1M2} at finite $M_2$ and $M_1=N\rho_1$ to get
for the distribution of $M_2$ at given $\rho_1$
\be
P(M_2,N|\rho_1) =
\frac{1}{2-\rho_1}\left(\frac{1-\rho_1}{2-\rho_1}\right)^{M_2} +
\frac{1-\rho_1}{2-\rho_1}\left(\frac{1}{2-\rho_1}\right)^{M_2}\ .
\label{PM2_fixed_rho1}
\ee
Multiplying by $z_2^{M_2}$, normalizing and taking the average of
$M_2$ then gives the result stated in\eq{minmax_mixed}. Note
that\eq{PM2_fixed_rho1} has a simple interpretation in the labelling
picture: For the $M_1=N\rho_1$ minima we have used up as many `left'
labels (and one `right' label). The probability that the largest
number $y_{2N}$ will be labelled `left' is then
$(N-N\rho_1)/(2N-N\rho_1-1)\to(1-\rho_1)/(2-\rho_1)$ for large $N$. The
first term in\eq{PM2_fixed_rho1} thus gives the probability that
$y_{2N}, y_{2N-1},\ldots,y_{2N-M_2+1}$ are all labelled `left' and the
next number down, $y_{2N-M_2}$, is labelled `right'; the second term
gives the analogous contribution from the reverse labelling. The
respective probabilities $(1-\rho_1)/(2-\rho_1)$ and $1/(2-\rho_1)$
for a `left' and `right' label remain the same throughout as we are
only labelling finitely many ($M_2+1$) numbers and so for large $N$
the fraction of labels in the bag of either kind only changes negligibly.

The second approach parallels more closely the one taken for the
minima problem. 
%
%
We first calculate the joint 
probability distribution, $P(\even{x}\i,\even{x}\a)$ of the smallest and 
largest number in the left box. The probability  that all the
$\even{x}_i$ are greater that some value
$\even{x}\i$ and smaller than some other value $\even{x}\a$ is ${\cal
  P}=(\even{x}\a-\even{x}\i)^N$. This is also the probability that the
minimum of these numbers is larger than $\even{x}\i$, and the maximum
smaller than $\even{x}\a$, so the joint probability density of the
minimum and maximum is obtained by differentiation as
\begin{eqnarray}
P(\even{x}\i,\even{x}\a) & = & -\frac{\partial^2{\cal P}}
{\partial \even{x}\i\,\partial \even{x}\a}
\ =\ N(N-1)(\even{x}\a - \even{x}\i)^{N-2}. 
\label{minmax_distr}
\end{eqnarray}
There is an analogous expression for $P(\odd{x}\i,\odd{x}\a)$. 
The remaining $N-2$ numbers in each box are then distributed uniformly
between $\even{x}\i$ and $\even{x}\a$, and $\odd{x}\i$ and
$\odd{x}\a$, respectively. 

One can now represent the probability of getting $M_1$ minima and
$M_2$ maxima in terms of averages over this distribution. As before we
assume that the minima are in the left box, \ie\
$\even{x}\i<\odd{x}\i$, and multiply the probability by a factor 2 to
cover the opposite case:
\bea
\fl \frac{1}{2}P(M_1,M_2,N)&=&\Biggl\langle 
\Theta(\odd{x}\i-\even{x}\i)\Theta(\even{x}\a-\odd{x}\a)
\frac{(N-2)!}{(M_2-1)!(N-M_1-M_2)!(M_1-1)!}
\nonumber\\
& &\times
\left(\frac{\even{x}\a-\odd{x}\a}{\even{x}\a-\even{x}\i}\right)^{M_2-1} 
\left(\frac{\odd{x}\a-\odd{x}\i}{\even{x}\a-\even{x}\i}\right)^{N-M_1-M_2}
\left(\frac{\odd{x}\i-\even{x}\i}{\even{x}\a-\even{x}\i}\right)^{M_1-1} 
\Biggr\rangle
\nonumber
\\
\fl & &{}+\Biggl\langle 
\Theta(\odd{x}\i-\even{x}\i)\Theta(\even{x}\a-\odd{x}\i)
\Theta(\odd{x}\a-\even{x}\a)
\nonumber\\
& &\times
\binom{N-2}{M_1-1}
\left(\frac{\even{x}\a-\odd{x}\i}{\even{x}\a-\even{x}\i}\right)^{N-M_1-1}
\left(\frac{\odd{x}\i-\even{x}\i}{\even{x}\a-\even{x}\i}\right)^{M_1-1} 
\nonumber\\
& &\times
\binom{N-2}{M_2-1}
\left(\frac{\odd{x}\a-\even{x}\a}{\odd{x}\a-\odd{x}\i}\right)^{M_2-1} 
\left(\frac{\even{x}\a-\odd{x}\i}{\odd{x}\a-\odd{x}\i}\right)^{N-M_2-1}
\Biggr\rangle
\nonumber
\\
\fl & &{}+\delta_{M_1,N}\delta_{M_2,N} \left\langle
\Theta(\odd{x}\i-\even{x}\i)\Theta(\odd{x}\a-\even{x}\a)
\Theta(\odd{x}\i-\even{x}\a)\right\rangle\ .
\label{PM1M2_integral}
\eea
The three terms on the r.h.s.\ are arranged in the same order as
in\eq{PM1M2}, and represent different orderings of $\even{x}\i$,
$\odd{x}\i$, $\even{x}\a$ and $\odd{x}\a$. In the first term, the
Theta functions and the constraint $\odd{x}\i<\odd{x}\a$ enforce the
ordering $\even{x}\i<\odd{x}\i<\odd{x}\a<\even{x}\a$, so that the left
box contains both minima and maxima and the right box neither. The
remaining factors in this term give the probability that out of the
$N-2$ numbers in the left box (other than $\even{x}\i$ and
$\even{x}\a$) exactly $M_1-1$ are below $\odd{x}\i$ and hence minima,
and $M_2-1$ are above $\odd{x}\a$ and therefore maxima. The second term
corresponds to the ordering
$\even{x}\i<\odd{x}\i<\even{x}\a<\odd{x}\a$, where the maxima are in
the right box but the ranges of numbers in the two boxes still
overlap. In this case we need to find $M_1-1$ numbers in the left box
(in addition to $\even{x}\i$) that are below $\odd{x}\i$, and $M_2-1$
numbers in the right box (in addition to $\odd{x}\a$) that are above
$\even{x}\a$. Finally, the last term is for the ordering
$\even{x}\i<\even{x}\a<\odd{x}\i<\odd{x}\a$. All numbers in the left
box are then smaller than in the right one, and we have $M_1=N$ minima
on the left and $M_2=N$ maxima on the right.

In the representation\eq{PM1M2_integral} one can easily perform the
sums defining the generating function\eq{Gz1z2} to get
\bea
\fl \frac{1}{2}G(z_1,z_2,N)&=&\Biggl\langle 
\Theta(\odd{x}\i-\even{x}\i)\Theta(\even{x}\a-\odd{x}\a)
\nonumber\\
& &\times 
z_1 z_2 \left(\frac
{z_2(\even{x}\a-\odd{x}\a)+\odd{x}\a-\odd{x}\i+z_1(\odd{x}\i-\even{x}\i)}
{\even{x}\a-\even{x}\i}\right)^{N-2}
\Biggr\rangle
\nonumber
\\
\fl & &{}+\Biggl\langle 
\Theta(\odd{x}\i-\even{x}\i)\Theta(\even{x}\a-\odd{x}\i)
\Theta(\odd{x}\a-\even{x}\a)
\nonumber\\
& &\times 
z_1 z_2 \left(\frac
{\even{x}\a-\odd{x}\i+z_1(\odd{x}\i-\even{x}\i)}
{\even{x}\a-\even{x}\i}
\right)^{N-2}
\left(\frac
{z_2(\odd{x}\a-\even{x}\a)+\even{x}\a-\odd{x}\i}
{\odd{x}\a-\odd{x}\i}
\right)^{N-2} \Biggr\rangle
\nonumber
\\
\fl & &{}+z_1^N z_2^N \left\langle
\Theta(\odd{x}\i-\even{x}\i)\Theta(\odd{x}\a-\even{x}\a)
\Theta(\odd{x}\i-\even{x}\a)\right\rangle\ .
\label{Gz1z2_integral}
\eea
To carry out the averages one inserts\eq{minmax_distr} and its
analogue for the right box and integrates over $\even{x}\i$,
$\odd{x}\i$, $\even{x}\a$ and $\odd{x}\a$. In each term two of the
integrals can be done directly and one is left with
\bea
\fl \frac{1}{2}G(z_1,z_2,N)&=&
\int_0^1 \!d\odd{x}\i \!\int_{\odd{x}\i}^1 \!d\odd{x}\a
N(N\!-\!1)(\odd{x}\a-\odd{x}\i)^{N-2}
\Bigl\{ 
[z_2(1-\odd{x}\a)+\odd{x}\a-\odd{x}\i+z_1\odd{x}\i]^N
\nonumber\\
\fl & &{}
-[z_2(1-\odd{x}\a)+\odd{x}\a-\odd{x}\i]^N
-[\odd{x}\a-\odd{x}\i+z_1\odd{x}\i]^N
+[\odd{x}\a-\odd{x}\i]^N\Bigr\}
\nonumber
\\
\fl & &{}+
\int_0^1 d\odd{x}\i \int_{\odd{x}\i}^1 d\even{x}\a\,
N^2 \left\{
[\even{x}\a-\odd{x}\i+z_1\odd{x}\i]^{N-1}
-[\even{x}\a-\odd{x}\i]^{N-1}
\right\}
\nonumber\\ \fl & &
\times \left\{
[z_2(1-\even{x}\a)+\even{x}\a-\odd{x}\i]^{N-1}
-[\even{x}\a-\odd{x}\i]^{N-1}
\right\}
\nonumber
\\
\fl & &{}+z_1^N z_2^N 
\int_0^1 d\even{x}\a \int_{\even{x}\a}^1 d\odd{x}\i\,
N^2 \even{x}\a^{N-1}(1-\odd{x}\i)^{N-1}\ .
\label{Gz1z2_simplified}
\eea
The remaining integrals in the last line can of course also be done and give
$[(2N)!/N!^2]^{-1}$ as expected from\eq{PM1M2}.

From here on one can proceed as in the minima-only case. If both $z_1$
and $z_2$ are below 2, one rescales $\odd{x}\i=u/N$, $\odd{x}\a=1-v/N$
in the first integral and similarly for the other terms; this gives
$G(z_1,z_2,N\to\infty)=[z_1/(2-z_1)][z_2/(2-z_2)]$ as derived by a
different route in the main text. For larger $z_1$ or $z_2$ one uses
steepest descents again. The functions to be maximized always have
negative Hessian determinants so the maxima are on the boundary.  We illustrate
only the mixed case $z_1>z_2$ at $1/z_1+1/z_2>1$. Here the relevant
saddle point in the first integral is $\odd{x}\i=(z_1-2)/[2(z_1-1)]$
and $\odd{x}\a=1$. Because this is at the upper extreme of the
integration range of $\odd{x}\a$, however, one needs to rescale
$\odd{x}\a=1-v/N$ to treat the near-cancellation of the first and
third terms in the integrand explicitly. (The second and fourth terms
make exponentially subleading contributions.) In the second integral
one needs to set similarly $\even{x}\a=1-v/N$ to capture the
near-cancellation in the second factor. One thus gets for large $N$,
after neglecting the exponentially subdominant third term
of\eq{Gz1z2_simplified}:
\bea
\fl \frac{1}{2}G(z_1,z_2,N)&=&
\int_0^1 \!d\odd{x}\i \!\int_0^\infty \!dv\,
(N\!-\!1)(1-\odd{x}\i)^{N-2}e^{-v/(1-\odd{x}\i)}
\nonumber
\\
\fl & &\times
[1+(z_1-1)\odd{x}\i]^N
\Bigl\{ e^{(z_2-1)v/[1+(z_1-1)\odd{x}\i]} -
e^{-v/[1+(z_1-1)\odd{x}\i]}
\Bigr\}
\nonumber
\\
\fl & &{}+
\int_0^1 d\odd{x}\i \int_0^\infty dv\,
N [1+(z_1-1)\odd{x}\i]^{N-1}
e^{-v/[1+(z_1-1)\odd{x}\i]}
\nonumber\\ \fl & &
\times (1-\odd{x}\i)^{N-1}
\left\{
e^{(z_2-1)v/(1-\odd{x}\i)} -
e^{-v/(1-\odd{x}\i)}
\right\}\ .
\eea
The common exponential
factor $\{(1-\odd{x}\i)[1+(z_1-1)\odd{x}\i]\}^N$ means 
that for large $N$ we can replace $\odd{x}\i=(z_1-2)/[2(z_1-1)]$ in
all other, slowly varying, terms to obtain
\bea
\fl \frac{1}{2}G(z_1,z_2,N)&=&
\frac{2N(z_1-1)^2z_2(z_2-2)}{z_1(z_1-z_2)(z_1 z_2 - z_1 -
  z_2)}
\int_0^1 d\odd{x}\i
\{(1-\odd{x}\i)[1+(z_1-1)\odd{x}\i]\}^{N}\ .
\eea
This result is asymptotically exact for $N\to\infty$. It depends on 
$z_2$ only through subexponential factors, which is why the
cancellations referred to above have to be treated so carefully.
Using $\langle M_2\rangle = \partial \ln G/\partial \ln z_2$
then retrieves after a little algebra the result for the number of maxima
in the mixed phase stated in\eq{minmax_mixed}.

\section{Beta function asymptotics}

Here we gather the asymptotic properties of the incomplete Beta
function $B(p,q;a)=\int_0^a dt\,t^{p-1}(1-t)^{q-1}$ that we need in
Sec.~\ref{sec:Bethe_min}. Specifically, setting $p=1/(\mu+1)$, we
require the behaviour of $B(p,p;a)$, $B(p,p+1;a)$ and $B(p+1,p;a)$ in
the limit $p\to 0$. Directly from the definition one sees that these
three functions are linked by the simple sum rule
\be
B(p,p;a)=B(p,p+1;a)+B(p+1,p;a)\ .
\label{beta_sum}
\ee
We will always keep $a<1$, with $a$ either fixed as $p\to 0$ or itself
going to zero.

The last function in\eq{beta_sum} is simplest as it remains
non-singular:
\be
B(p+1,p;a) = \int_0^a dt\,t^{p}(1-t)^{p-1} \to \int_0^a dt\,(1-t)^{-1}
= -\ln(1-a)\ .
\label{B_pplus1_p}
\ee
The other two functions, on the other hand, diverge as $p\to 0$. With
the variable transformation $s=t^{p}$ one gets
\be
p B(p,p;a) = p\int_0^a dt\,t^{p-1}(1-t)^{p-1} 
= \int_0^{a^p} ds\,\left(1-s^{1/p}\right)^{p-1}\ .
\ee
Now if $a$ vanishes quickly enough (exponentially in $1/p$) when $p\to
0$ for $a^p$ to stay bounded below 1, we can exploit the fact that the
integrand approaches unity for all $s<1$ to get
\be
p B(p,p;a) \to a^p\ .
\label{Beta_exponential_a}
\ee
From\eq{beta_sum} and\eq{B_pplus1_p}, which shows that $B(p+1,p;a)$ stays
finite, the same limit applies to $B(p,p+1;a)$.

The result\eq{Beta_exponential_a} does in fact extend also to $a$ that
vanish more slowly or stay finite, so that $a^p\to 1$. One can see
this by subtracting off the leading term:
\be
B(p,p;a)-\frac{1}{p}a^p = \int_0^a dt\,t^{p-1}[(1-t)^{p-1}-1]\ .
\ee
The integrand now remains non-singular for $p\to 0$ and approaches
$1/(1-t)$, so that
\be
B(p,p;a)-\frac{1}{p}a^p \to -\ln(1-a)\ .
\ee
Multiplying by $p$ gives back\eq{Beta_exponential_a} as claimed. For fixed
$a$ it is more useful to rewrite the last relation, using $a^p=1+p \ln a +
\order(p^2)$, as
\be
B(p,p;a)-\frac{1}{p} \to \ln[a/(1-a)]\ .
\label{Beta_order1}
\ee

\end{document}